\documentclass{aa}  

\usepackage{graphicx}

\usepackage{txfonts}
\usepackage[colorlinks=true,linkcolor=blue,citecolor=blue]{hyperref}

\begin{document}

   \title{Properties of intra-cluster low-mass X-ray binaries in Fornax globular clusters }
   \subtitle{}

   \author{G. Riccio\inst{1} \and M. Paolillo\inst{1,2,3} \and M. Cantiello\inst{4} \and R. D'Abrusco\inst{5} \and X. Jin\inst{6} \and Z. Li\inst{7} \and T. Puzia\inst{8} \and S. Mieske\inst{9} \and D.J. Prole\inst{10,11} \and E. Iodice\inst{2} \and G. D'Ago\inst{8} \and M. Gatto\inst{1,3} \and M. Spavone\inst{2}}

   \institute{Department of Physics, University of Naples Federico II, C.U. Monte Sant'Angelo, Via Cinthia, I-80126 Naples, Italy
   \and
   INFN - Sezione di Napoli, 80126 Naples, Italy
   \and
   INAF-Osservatorio astronomico di Capodimonte, Via Moiariello 16, I-30131 Naples, Italy
   \and
   INAF-Oss.dell'Abruzzo, 64100 Teramo, Italy
   \and
   Harvard-Smithsonian Center for Astrophysics, Cambridge, MA 02138, USA 
   \and
   Steward Observatory, University of Arizona, 933 N. Cherry Ave., Tucson, AZ 85719, USA.
   \and
   School of Astronomy and Space Science, Nanjing University, China
   \and
   Institute of Astrophysics, Pontificia Universidad Católica de Chile, Av. Vicuña Mackenna 4860, 7820436 Macul, Santiago, Chile
   \and
   European Southern Observatory, Santiago, Chile 
   \and 
   Research Centre for Astronomy, Astrophysics \& Astrophotonics, Macquarie University, Sydney, NSW 2109, Australia
   \and 
   Department of Physics \& Astronomy, Macquarie University, Sydney, NSW 2109, Australia
   }

 
  \abstract
  {} 
   {We present a study of the intra-cluster population of low-mass X-ray binaries (LMXB) residing in globular clusters (GC) in the central 1 $deg^2$ of the Fornax galaxy cluster. 
   Differently from previous studies, which were restricted to the innermost regions of individual galaxies, this work is aimed at comparing the properties of the intra-cluster population of GC-LMXBs with those of  the host galaxy. }
   {The data used are a combination of \textit{VLT Survey Telescope} (VST) and \textit{Chandra} observations.We perform a cross-match between the optical and X-ray catalogue, in order to identify the LMXBs residing in GCs. We divide the GC-LMXBs into host-galaxy and intra-cluster objects based on their distance from the nearest galaxy in terms of effective radius ($R_{eff}$). We found  82  intra-cluster  GC-LMXBs  and  86  objects that are hosted in galaxies. As the formation of LMXBs also depends on the host GC colour, we performed a Gaussian mixture model to divide the population into red and blue GCs.  }
   {As has been found for the innermost regions of galaxies, LMXBs tend to form in red and bright GCs in intra-cluster space as well. We find, however, that the likelihood of a red GC to host an LMXB decreases with galactocentric distance, but it remains approximately constant for the blue GC population. Investigating the X-ray properties of the LMXBs residing in GCs, we find a difference in the X-ray luminosity function between the intra-cluster and host-galaxy sample: both follow a power-law down to $\sim 8.5\times 10^{37}$ erg s$^{-1}$, which is consistent with field LMXBs for the intra-cluster sample, while the latter agree with previous estimates for LMXBs in GCs. We observe a deficiency of bright LMXBs in blue intra-cluster GCs, however. This might indicate a lack of black hole binaries in metal-poor systems. We further investigated the spectral properties of the GC-LMXBs through their hardness-ratio. We detect a tentative difference in the hardness ratio of two populations, where the intra-cluster GC-LMXBs appear to have harder spectra than the host-galaxy objects. We find the same trend when we compare red and blue GC-LMXBs: the spectra of the blue sample are harder spectra than those of the red sample. This result could suggest a relation between the spectral properties of LMXBs and the host GC colour and therefore its metallicity. 
   
   We discuss the possibilities of spatial biases due to uncertainties in the X-ray spectral response correction and due to contamination by background active galactic nuclei.} 
   
   {}

    
   \maketitle
    
\section{Introduction}
\label{sec:Introduction}
Low-mass X-ray binaries (LMXBs) are stellar systems composed of an extremely dense object (a neutron star or black hole) that accretes mass from a secondary star (a main-sequence star of about one solar mass). They represent the dominant X-ray binary (XRB) population in early-type galaxies. It has been shown that a significant fraction of LMXBs resides in GCs. This varies from 10\%-20\% in small galaxies and reaches $\sim 70$\% in cD galaxies, depending on the morphological type of the galaxy and on the specific abundance of the GCs \cite[e.g.][]{Kim2009}.

The GC-LMXB association is particularly interesting as the high stellar density near the centre of GCs may trigger the formation of binaries either by three-body process or by tidal capture. It was observed that LMXBs tend to form in bright GCs, as expected if the luminosity is a proxy for the total number of stars they contain (\citealt{Fabbiano2006}, and references therein). On the other hand, size and concentration reflect the efficiency of dynamical interaction and favour binary formation in dense environments. Furthermore, the formation can also be influenced by the mass, size, and metallicity of GCs. Red metal-rich GCs are $\text{about three}$ times more likely to host LMXB than blue (metal-poor) GCs, in part because red GCs are denser on average than the blue counterpart (\citealt{Jordan2004}, \citealt{Fabbiano2006}, \citealt{Paolillo2011}, \citealt{D'Ago2014} and references therein). However, the role played by metallicity in LMXB formation is still unclear, and understanding this connection would help us to know how these objects are formed, and if the properties of the environment (the host GC or the galaxy properties) can have an impact on the structure and emission of the LMXB.

The fact that the spatial distribution of GC-LMXBs in early-type galaxies usually follows the distribution of the host GC population well (\citealt{Paolillo2011}) suggests that the likelihood of LMXB formation is mainly driven by the internal GC properties. However, there have been claims that LMXBs may be less concentrated than GCs around giant ellipticals, suggesting that environmental effects may influence the formation and evolution of LMXBs. Unfortunately, most studies so far were limited to the central regions of galaxies because the area explored by past surveys (\textit{Hubble Space Telescope}) was limited and high-resolution imaging was lacking. This imaging is needed to select and measure the structural parameters of extragalactic GCs. This has prevented studies of the effect of the distance from the galaxy centre on the formation of LMXBs inside GCs. Because the spatial distribution of the red and blue GCs changes according to the distance from the galaxy, moreover, wide-field observations are required to distinguish the different physical
processes at work in a scenario in which red GCs are associated with the main body of the galaxy while the blue GCs are associated with the halo (\citealt{Cantiello2018}).

Significant populations of intra-cluster GCs have been discovered in the Virgo (\citealt{Durrell2014}) and in the Fornax galaxy cluster. Using the \textit{VLT Survey Telescope} wide-field imaging obtained within the Fornax Deep Survey (FDS, \citealt{Iodice2016}) in the Fornax cluster, our collaboration has proved the existence of a vast population of intra-cluster stellar systems extending out to a significant fraction of the viral radius (\citealt{D'Abrusco2016}; \citealt{Cantiello2018}; \citealt{Cantiello2020}). This was found to match the distribution of the intra-cluster light (\citealt{Iodice2017}), thus tracing the past dynamical evolution of the galaxy cluster itself. In addition, \cite{Jin2019} discovered a population of field LMXBs throughout the cluster, around the dominant cD galaxy NGC1399, by analysing \textit{Chandra} archival images.

In this work, we investigate the properties of LMXBs residing in this extended population of intra-cluster GCs making use of the latest FDS data release, combined with \textit{Chandra} X-ray data covering the core of the Fornax galaxy cluster. Based on this, we study the GC-LMXB connection within the whole Fornax cluster and its dependence on host galaxy, environment, galactocentric distance, and metallicity.

\section{Dataset}
This work is based on the combination of VST and \textit{Chandra} data. We briefly summarize the main properties of the datasets. We refer to the cited papers for more details.

\subsection{Optical data}
\label{optical_data}

The optical data used were acquired as part of the  Fornax FDS  based on observations obtained in u, g, r,  and i bands with the Survey Telescope (VST) of the Very Large Telescope at the ESO Paranal Observatory. The VST is a wide-field optical imaging telescope with a 2.6-meter aperture with a field of view (FoV) of 1 $degree^2$. The telescope is equipped with the 268 megapixel OmegaCAM  with a pixel scale of $0''.21$ $pixel^{-1}$. The survey was designed to map the entire Fornax cluster out to the virial radius, and it also covered the NGC1316 subgroup \citep{Iodice2017}. Located at a distance of $D=20.13\pm0.4$ $Mpc$ \citep{Blakeslee2009}, this cluster is the second nearest to us after the Virgo cluster and therefore represents an ideal target for this type of study.
The survey observed $\sim 27$ square degrees approximately centred on NGC1399 and NGC1316 (\citealt{Cantiello2020}), reaching magnitude limits for point-like sources of 24.1, 25.2, 24.6, and 23.6 in the \textit{u, g, r, \textup{and} i} bands, respectively (AB mag photometric system), while the median seeing of the observation ranges from 0.6 to 1.1 arcsec. We limit our analysis to the central 1.5 square degrees around NGC1399, which overlaps with the \textit{Chandra} X-ray coverage of the cluster core.  The data reduction was performed using the Astro-WISE pipeline \cite[see e.g.][]{Venhola2019}. This is a tool for the reduction of large field data to perform pre-reduction (subtraction of bias, correction of flat), illumination, and edge correction and photometric and astrometric calibrations. To properly detect and study GCs, we need to minimize the contamination from the brightest galaxies in the cluster (e.g. NGC1399 and NGC1404). In order to model and subtract the galaxies, ELLIPSE task in IRAF STSDAS is used during the catalogue preparation (\citealt{Jedrzejewski1987}). To produce a complete catalogue of all sources present in the VST field of view, a combination of SExtractor (\citealt{Bertin}) and DAOphot \citep{Stetson1984} was used on the galaxy-subtracted frame independently in each filter. For additional information on the data reduction and source photometry, we refer to \citealt{Cantiello2020}.  

\begin{figure*}[t]
    \centering
    \includegraphics[width=1\hsize]{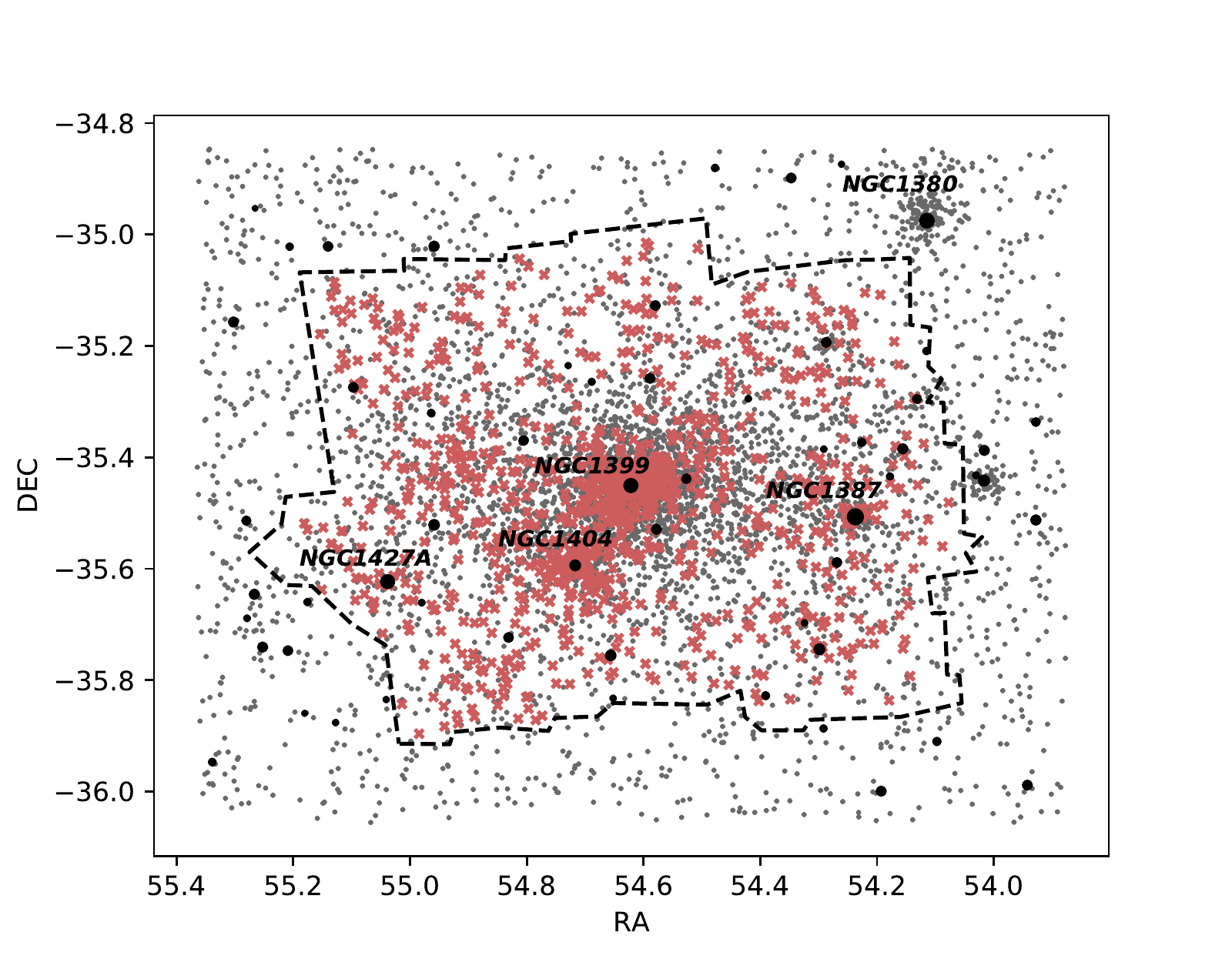}
    \caption{Spatial distribution of the GCs (grey points) and the X-ray sources (red crosses) centred on NGC1399. The solid black circles represent bright galaxies in the cluster from \cite{Ferguson1998}. The size is proportional to their effective radii. The dashed line highlights the FoV of \textit{Chandra} observations.}
    \label{fig:d'abrusco}
\end{figure*}

At the Fornax distance, GCs appear unresolved from the ground. They are therefore hard to separate from stars and compact background galaxies. The selection of the GC sample accordingly represents a crucial step in our analysis. In this case, we adopted the approach presented in \cite{Cantiello2018} applied to the improved FDS dataset (\citealt{Cantiello2020}). The GC selection criteria were defined using the training set of spectroscopically confirmed GCs published in  \cite{Pota2018} and \cite{Schuberth2010}. We refer to figure 1 of \cite{Pota2018} for the region covered by the confirmed GCs. 

In order to minimize the contamination from foreground sources, we further discarded all sources with a \textit{g} magnitude brighter than $M_{TO}-3\sigma_{GCLF}$, where $M_{TO}$ is defined as the absolute magnitude turn-off of the Gaussian GC luminosity distribution, assuming $\sigma_{GCLF}=1.4\pm0.1$ $mag$ and $M_{TO}=-7.4\pm0.2$ $mag$ \citep[e.g.][]{Villegas2010}. At the distance of Fornax, the turn-over magnitude corresponds to $m_{TO}\sim24.0$ $mag$. The adopted selection criteria are given in Table \ref{table:1}.

\begin{table}
\caption{Selection criteria for the GC candidates.}   
\label{table:1}      
\centering                          
\begin{tabular}{c c}        
\hline\hline                 
Parameters & Values \\    
\hline                        
   \textbf{magnitude} & \textbf{$19.8\leq m_g \leq 26$} \\      
   \textbf{concentration index$^1$} & $0.8 \leq CI_n \leq 1.15$ \\
  \textbf{colour} &$0.6 \leq g-i \leq 1.45$ \\
 & $1.35 \leq u-r \leq 3.5$ \\
  \textbf{Difference from model$^2$} & $\leq 0.4$ \\
\hline                                   
\end{tabular}
\vspace{1mm}
 \scriptsize{
 \\
 {\it Notes:} $^1$\textit{$CI_n$}: normalized concentration index based on the difference in \textit{g}-band magnitude between apertures of 6 and 12 pixels. 
  $^2$Maximum distance from the best-fit population synthesis model for spectroscopically confirmed GCs in the colour-colour diagram.}
\end{table}

The catalogue of candidate GCs contains 5178 sources. Their spatial distribution is shown in Figure \ref{fig:d'abrusco}.  The figure shows the extended distribution of GCs that was reported by \cite{D'Abrusco2016}. While a large fraction of these GCs are clustered around the brightest galaxies, many GCs occupy the intra-cluster space, several effective radii ($R_{eff}$) away from any cluster galaxy member. The positions of cluster galaxies and their effective radii are extracted from \cite{Ferguson1998}. In Table \ref{table:2} we show the $R_{eff}$ of the galaxies in the FoV of the \textit{Chandra} observations. We point out that the $R_{eff}$ of NGC1399 and NGC1404 are considerably smaller than those reported in \cite{Iodice2016} and \cite{Iodice2019}. These values were estimated using very deep observations probing very low surface brightnesses where the extended stellar halos of NGC1399 (and NGC1404) merge with the intra-cluster light and encompass most of the Fornax cluster. To identify intra-cluster sources, we therefore preferred to use the value provided by \cite{Ferguson1998}.

\begin{table}
\small
\caption{ List of galaxies from \cite{Ferguson1998} in the \textit{Chandra} FoV.}  
\label{table:2}

\begin{tabular}{l l l r}
\hline\hline
  \multicolumn{1}{c}{Name} &
  \multicolumn{1}{c}{RA ($deg$)} &
  \multicolumn{1}{c}{DEC ($deg$)} &
  \multicolumn{1}{c}{$R_{eff}$ ($kpc$)} \\ [0.5ex] 
\hline\\

  NGC1387 & 54.2370 & -35.5066 & 4.86\\
  NGC1399 & 54.6216666 & -35.45055 & 3.61\\
  NGC1427A & 55.0383333 & -35.62305 & 3.52\\
  NGC1389 & 54.29875 & -35.74444 & 1.95\\
  NGC1404 & 54.7170833 & -35.59388 & 1.94\\
  NG115 & 54.1558333 & -35.38472 & 1.74\\
  NG6 & 54.83125 & -35.72333 & 1.61\\
  NGC1380B & 54.2870833 & -35.19361 & 1.58\\
  NG5 & 55.0970833 & -35.27444 & 1.54\\
  G72& 54.8054166 & -35.36972 & 1.40\\
  NGC1381 & 54.1316666 & -35.29527 & 1.25\\
  NG8 & 54.26875 & -35.58861 & 1.17\\
  NG21& 54.5779166 & -35.52916 & 1.14\\
  G79& 54.22625 & -35.37305 & 1.10\\
  NG23 & 54.3908333 & -35.82777 & 0.99\\
  NGC1396 & 54.5266666 & -35.43833 & 0.95\\
  NG87 & 54.96375 & -35.32055 & 0.89\\
  NG7 & 54.5795833 & -35.1275 & 0.82\\
  NG116 & 54.1779166 & -35.43416 & 0.75\\
  NG72 & 54.68875 & -35.26444 &  0.74\\
  NG117 & 54.65625 & -35.75555 & 0.72\\
  NG111 & 54.98 & -35.66083 & 0.66\\
  D117 & 54.3241666 & -35.69722 & 0.62\\
  FCC197 & 54.29125 & -35.38527 & 0.61\\
  NG114 & 54.4204166 & -35.29472 & 0.61\\
  NG118 & 54.6520833 & -35.8325 & 0.60\\
  NG71 & 54.58875 & -35.25833 & 0.54\\
  D138 & 54.7291666 & -35.23527 & 0.54\\
\hline
\end{tabular}
\end{table}

\begin{figure}
    \centering
    \includegraphics[width=\hsize]{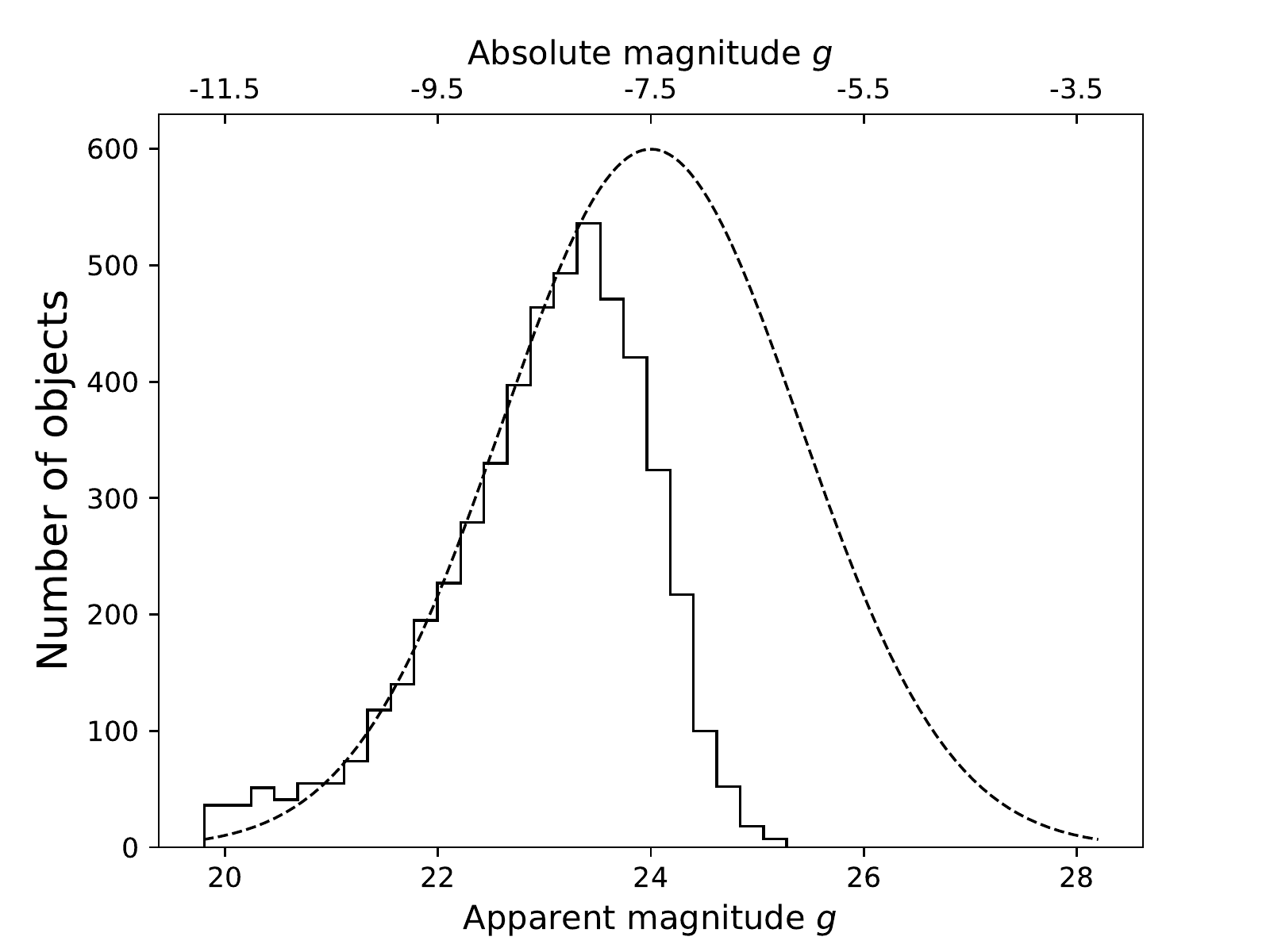}
    \caption{ Luminosity function in g band for the sample of candidate GCs. The dashed black line corresponds to the expected theoretical GC luminosity function (with values $\mu=24$ and  $\sigma=1.4$).}
    \label{fig:g_lfunc}
\end{figure}

The completeness of the original \textit{gri} catalogue is $\sim 80\%$ for magnitudes $g\leq24$ mag. However, the selection criteria introduced above in order to minimize the contamination of foreground/background sources combined with the use of the shallower \textit{u} band constrain the actual completeness of our sample, as shown in Fig. \ref{fig:g_lfunc}. We note that because LMXBs tend to reside in bright GC, the final detection limit of $m_g\sim 25$ does not represent a significant limitation for our study.

\begin{figure}
    \centering
    \includegraphics[width=\hsize]{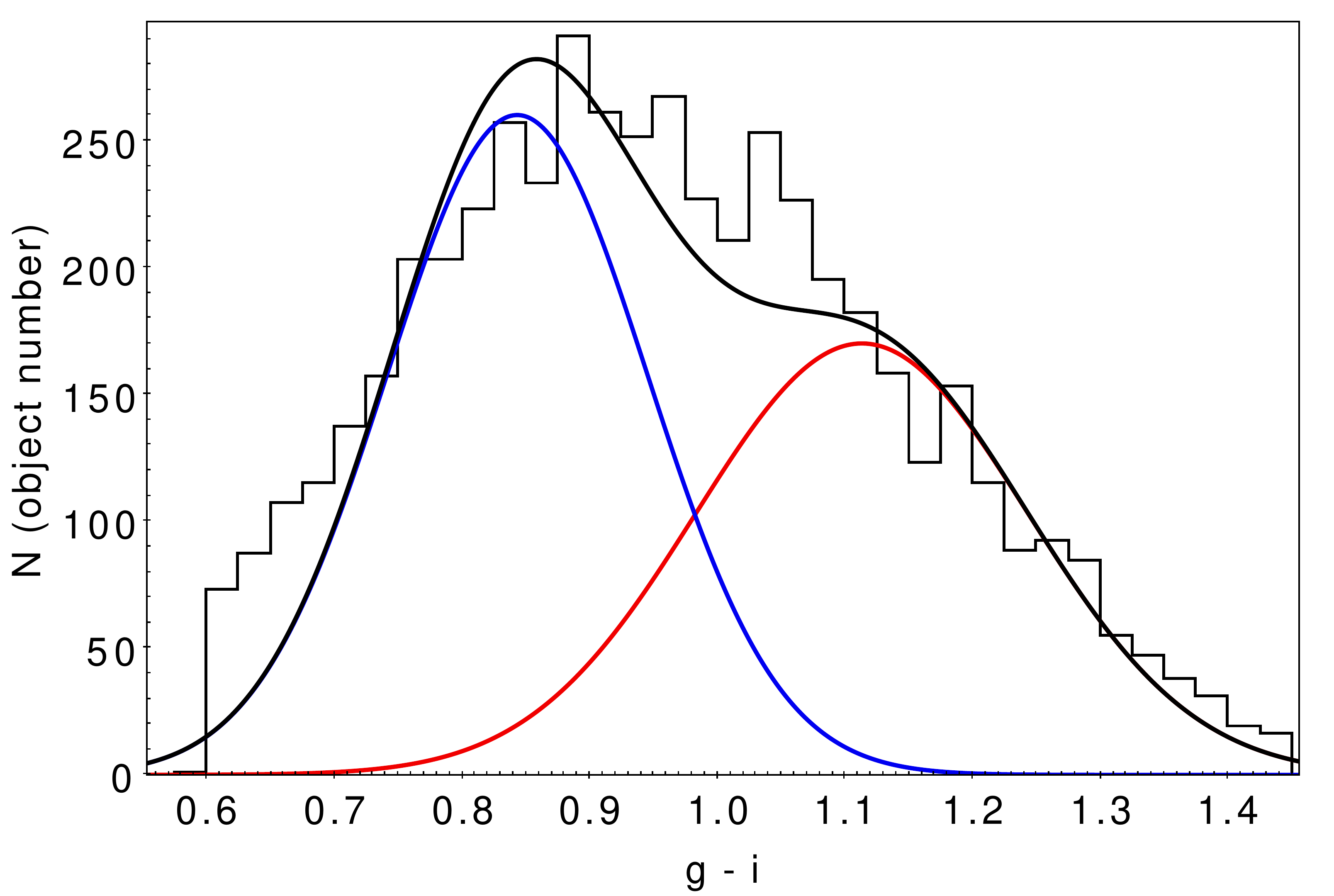}
    \caption{$g-i$ colour distribution of GC candidates in the central square degree of the Fornax cluster. The smooth solid  black, red, and blue lines show the GMM model for the total, red, and blue populations.}
    \label{fig:GMM}
\end{figure}

The colour distribution of GCs in the core of the Fornax cluster is known to be bimodal, mainly due to metallicity effects \citep[see e.g.][and references therein]{D'Abrusco2016,Cantiello2018}. In order to study the possible differences between LMXBs formed in different environments,  we used a \textit{Gaussian mixture model} (GMM; \citealt{Muratov2010}) to divide the whole population into red and blue GCs. Using a $g-i\sim1.00$ threshold, we find 2085 red GCs ($\sim$ 40\% of the total sample) and 3093 blue GCs ($\sim $60\% of the total sample). As discussed in more detail below, the relevance of each population is dependent on several factors, including the host galaxy properties, the galactocentric distance of the cluster, and the environment. We adopted a fixed colour threshold as we intend to study the overall intra-cluster population, independently of the individual galaxies. We therefore ignored for the moment the variations in the GC colour distribution as a function of the distance from the centres of the host galaxies (\citealt{Kim2013}; \citealt{Cantiello2015}) and the differences of the red/blue bimodality of the individual GCs (\citealt{Jordan2015}). In figure \ref{fig:GMM} we plot the GMM model over the g-i colour distribution for the entire sample of candidate GCs.  We point out that on cluster scales, the blue component dominates 3:2 of the total GC population. This is different from what has been reported close to individual galaxies, where red GCs are dominant \citep[e.g.][]{Puzia2014}. This agrees with the well-known trend for red and blue GC densities to have different radial gradients, the latter of which are shallower than the former. We note that the bulk of the blue GC population is redder than what was found for Fornax dwarf galaxies in any case \citep{Prole2019}.

\begin{figure}
    \centering
    \includegraphics[width=\hsize]{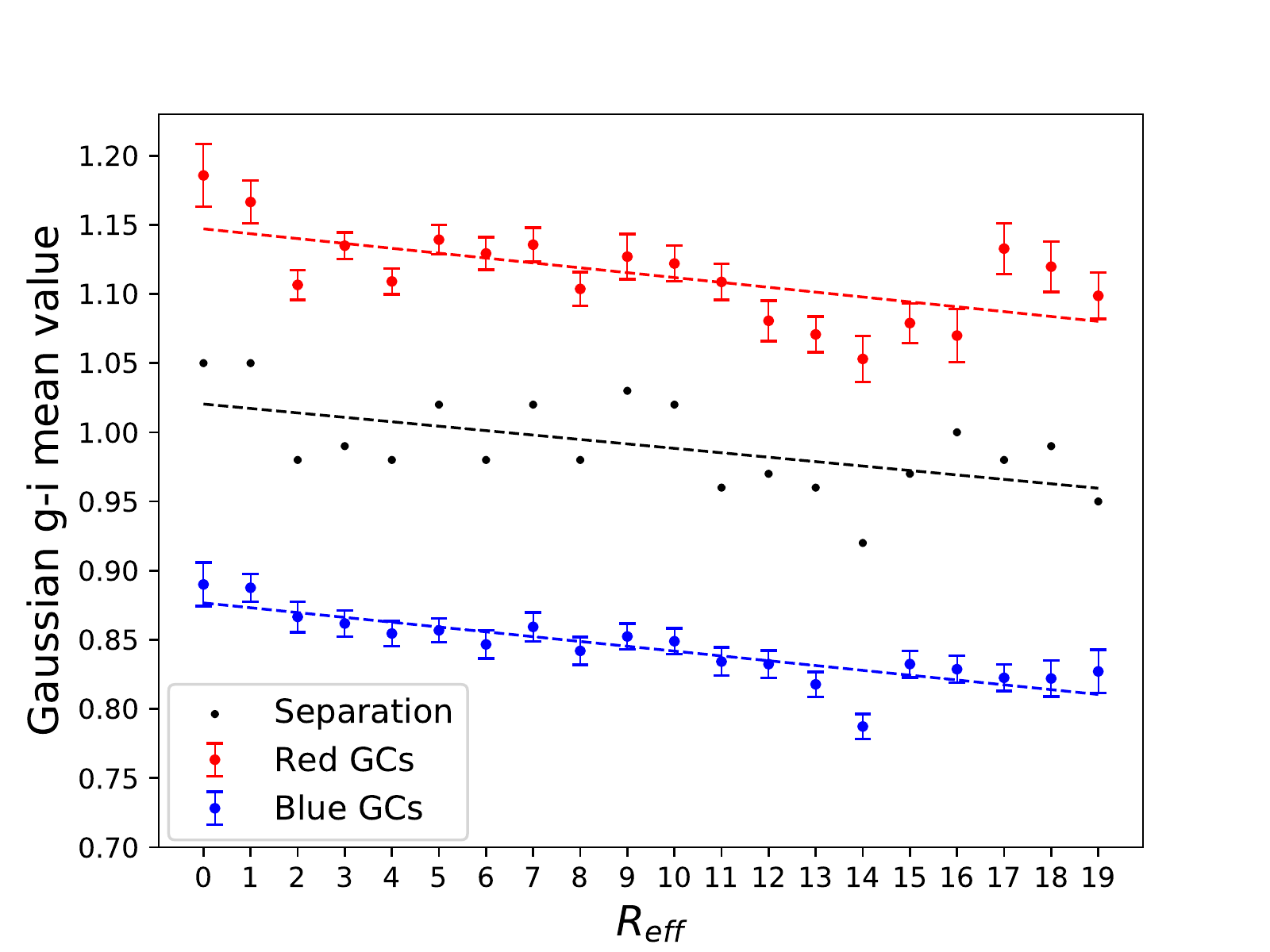}
    \caption{Mean values calculated by the GMM for different distances from each galaxy in terms of $R_{eff}$ for red and blue GCs, and the separation performed by the GMM. The error bars correspond to the standard deviation of the mean.}
    \label{fig:GMMdifference}
\end{figure}

However, we tried to quantify the variation in the colour distribution as a function of the distance by performing a GMM fit for different distances from each galaxy in terms of effective radius. We find a slight shift in the average value of each distribution towards more blue colours in progressively outer regions (Fig.\ref{fig:GMMdifference}). The slope of the fit is found to be $-0.0037 \pm 0.0005$ for the blue sample and $-0.0037 \pm 0.0017$ for the red sample. This is inconsistent/marginally consistent ($\sim 2.2\sigma$), respectively, with a constant trend. 
This means that not only does the intra-cluster space host a larger fraction of blue GCs, but also that both distributions gradually become bluer with increasing distance. For this result, we do not expect any systematic effect from background galaxy light because it was removed during the catalogue reduction process \citep{Cantiello2020}.

\subsection{X-ray data}
\label{Xray_data}
\label{sec:2.2}
 The X-ray data were extracted from 29 archival \textit{Chandra} observations (given in Tab. 1 of \citealt{Jin2019}) that were obtained with the \textit{Advanced CCD Imaging Spectrometer} (ACIS) for a total exposure time of 1.3 Ms. This covers large part of the central region of the Fornax cluster (Fig. \ref{fig:d'abrusco}). The details of the source detection procedure can be found in \cite{Jin2019}. We briefly summarize the procedure here. The X-ray data were analysed using the \textit{Chandra} Interactive Analysis of Observation (CIAO) tool \textit{Wavdetect}, producing counts and exposure maps, at the native pixel scale of $0.49\arcsec$, in three different bands: 0.5–2 (S band), 2–8 (H band), and 0.5–8 (F band) KeV. The exposure maps were weighted by an assumed incident spectrum of an absorbed power law with a photon index of 1.7 and an absorption column density $N_H=10^{21}$ $cm^{-2}$. This latter value is higher than the Galactic foreground absorption column ($\sim 1.5 \times 10^{20}$ $cm^{-2}$), but takes some intrinsic absorption of the LMXBs into account (see \citealt{Jin2019})

The centroids of the sources were refined using a maximum likelihood method that iterates over the positions of the individual counts within the 90\% of the \textit{enclosed counts radius} (ECR). The photon fluxes in various bands were calculated using a circular opening of 90\% of the ECR in non-crowded zones and 50\% for the crowded zones to avoid contamination between nearby sources. The position uncertainty (PU) at the 68\% confidence level was estimated following the empirical relation between PU, source counts, and source
position in terms of the off-axis angle.
At this point, a catalogue of 1279 independent sources was obtained, of which 1177 are in F band (0.5-8 KeV), 924 in S band (0.5-2 KeV), and 713 in H band (2-8 KeV). To derive the unabsorbed energy flux in the F band, a photon-to-flux conversion of $3.64\times 10^{-9}$ $erg/ph$ was used, assuming the previously cited incident source spectrum.

\begin{figure}
    \centering
    \includegraphics[width=\hsize]{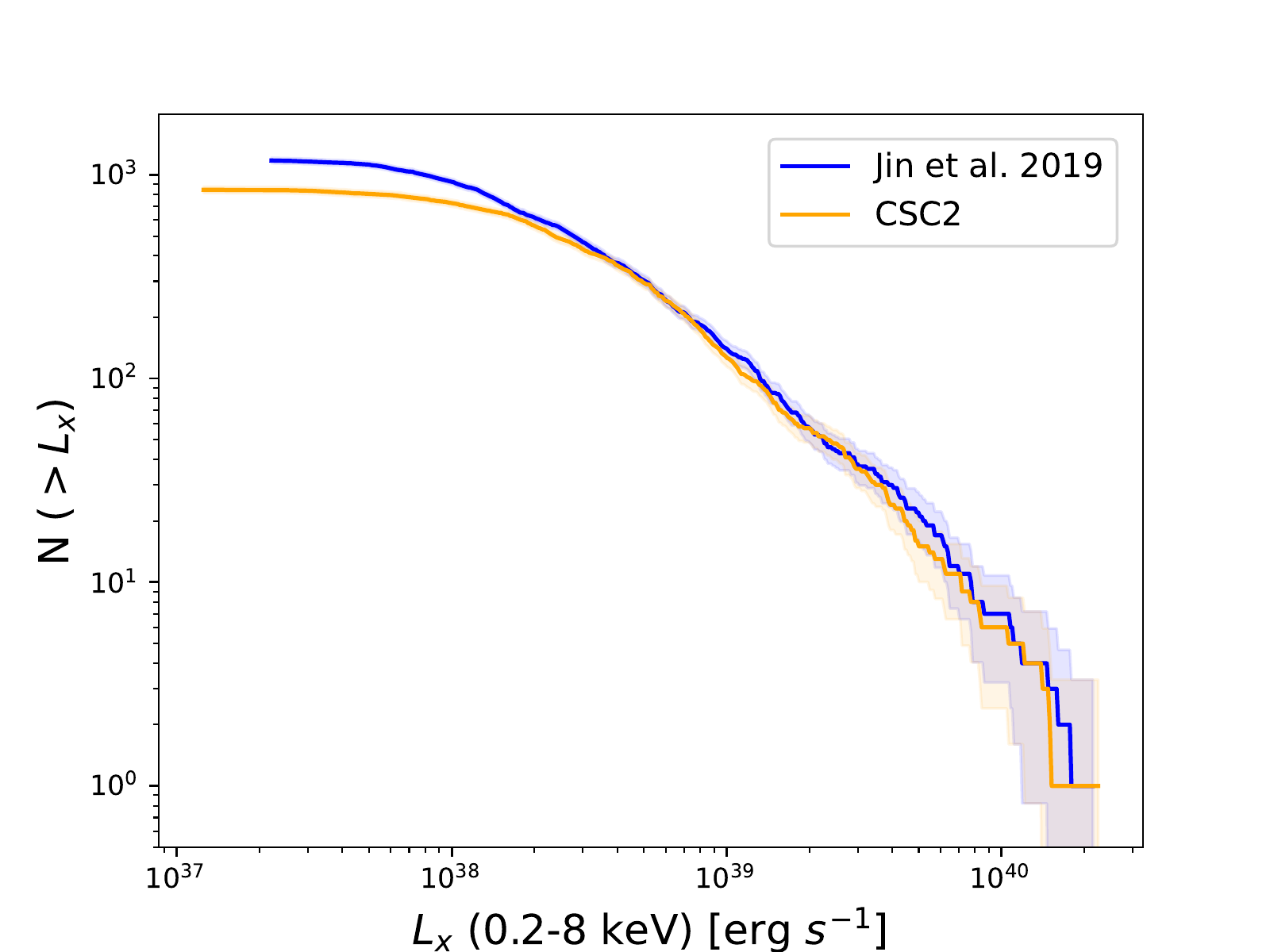}
    \caption{Cumulative X-ray luminosity function for the sources from the catalogue of \cite{Jin2019} and for the CSC catalogue. The shaded region represents the $1\sigma$ error.}
    \label{fig:compareX}
\end{figure}

In order to verify the quality and completeness of the catalogue, we compared it to the \textit{Chandra Source Catalog 2.0} (CSC master catalogue, http://cxc.harvard.edu/csc/). The CSC uses a different approach to compute the source fluxes (based on the individual photon energy instead of assuming an average conversion factor) and the exposure maps. In order to compare the two catalogues, we therefore cross-matched the source positions and compared the fluxes of the sources in common, deriving the rescaling factor ($r=1.41$) from CSC to \cite{Jin2019}. 
In figure \ref{fig:compareX} we show the cumulative X-ray luminosity function of the two catalogues after correcting the CSC fluxes as explained above. Both catalogues follow a similar truncated power-law distribution down to $LF\simeq 3\times10^{38}$ $erg/s$, after which 
the CSC LF flattens more rapidly than the Jin catalogue. The higher completeness of the Jin catalogue is due to two main factors: first, the CSC contains observations up to 2014, while the catalogue of \cite{Jin2019}  is based on all available \textit{Chandra} archived observations up to 2015. Of the 29 observations used in \cite{Jin2019}, the CSC lacks the two identified with ObsIDs 17549 and 14529, which are centreed on NGC1399 and NGC1404, where a large fraction of the GC population resides. Second, the CSC reports source properties in master, stack, and per-observation tables. A single source might then not be included in the master catalogue (used here) but might have entries at the stack or observation level. We verified that when we include sources from all tables, we obtain a comparable number of sources between the CSC and Jin catalogue (1004 and 1177, respectively), and this considerably reduces the disagreement. However, in this case, the choice for the measured properties that are used becomes complex (see the detailed discussion at https://cxc.cfa.harvard.edu/csc/organization.html).
For this reason, we mainly used the \cite{Jin2019} source catalogue and reverted to the CSC for specific measurements and tests. Most of our result would be confirmed, although with lower significance, using the CSC catalogue in any case, except for the hardness-ratio differences discussed in Sec.\ref{sec:4.2}

 \section{Identification of X-ray sources in GCs}
 \label{Sec:3}
In order to study the population of LMXBs residing in GCs, we performed a cross-match between the optical and the X-ray catalogues. To minimize the possibility of random matches, we employed the PU provided in the \cite{Jin2019} catalogue as matching radius for the X-ray
sources. This is different for each object. For the optical sources, we chose a matching radius of 0.5 arcsec that is typical of the VST astrometric accuracy (see e.g. \citealt{Capaccioli2015}). Two sources were considered to match when the separation between their positions did not exceed the sum of the two matching radii for the corresponding source. Based on the optical and X-ray source densities, we expect three false matches over the entire FoV. In addition, compact background galaxies hosting an active galactic nucleus (AGN) might be mistaken for GCs hosting an LMXB. This type of contamination is discussed in section \ref{sec:4.2}.
We identify 168 X-ray sources that are positionally coincident with GCs. We refer to these objects as GC-LMXBs.
 
 To study the photometric properties of the intra-cluster population, we divided the GC-LMXBs sources into \textit{host-galaxy} and \textit{intra-cluster} objects based on their projected distance from the nearest galaxy in terms of $R_{eff}$. 
 The choice on where to place the separation between the regions associated with the galaxy and the intra-cluster space is somewhat arbitrary. 

\begin{figure*}
    \centering
    \includegraphics[width=0.8\hsize]{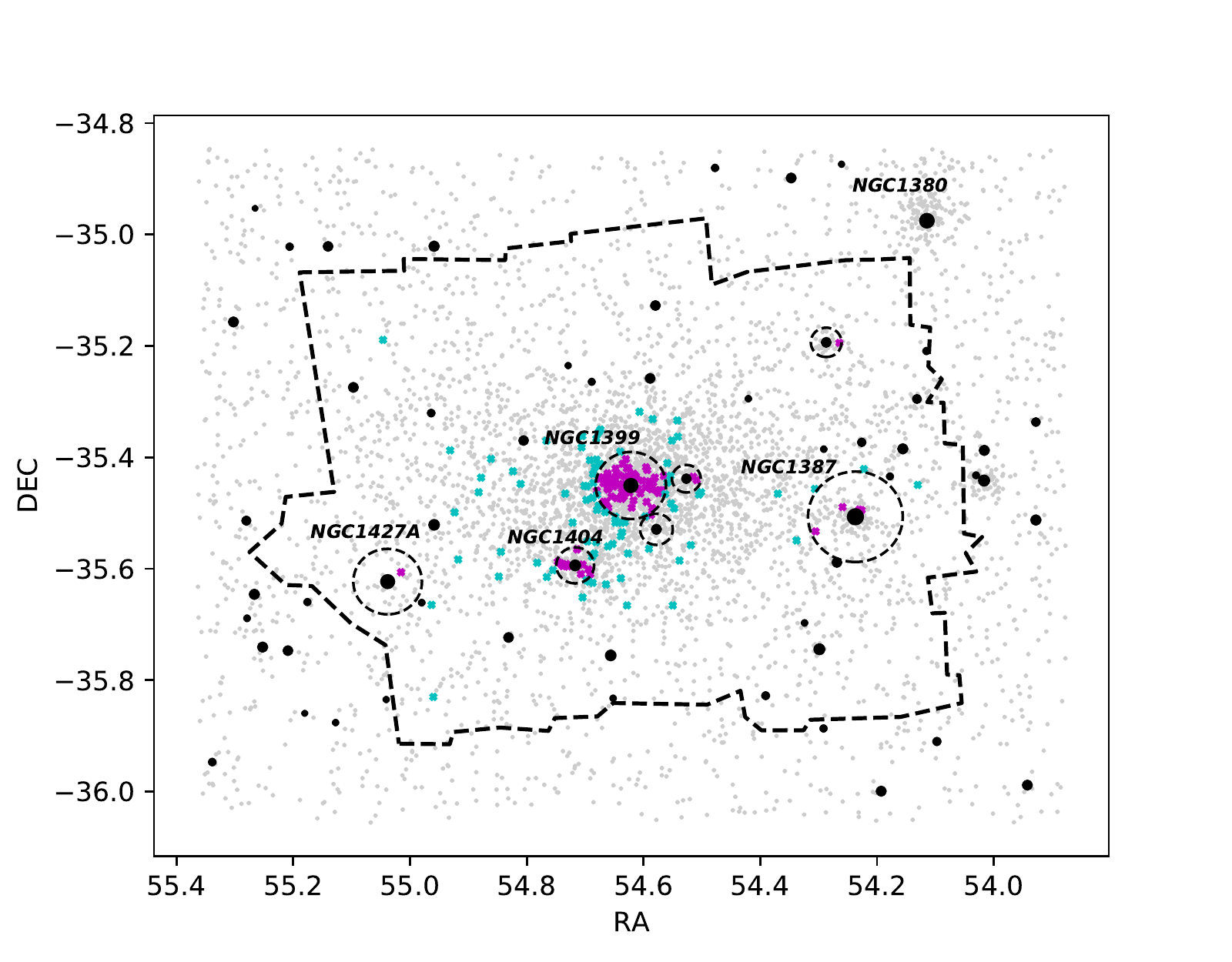}
    \caption{Spatial distribution of intra-cluster (cyan crosses) and host-galaxy (magenta crosses) GC-LMXBs. Cluster galaxies are the solid black circles, and the sizes are proportional to their effective radii. The dashed circle shows the 6 $R_{eff}$ distance from the galaxy centre. The dashed line depicts the FoV of the \textit{Chandra} observations.}
    \label{fig:SD}
\end{figure*}

Several authors considered 5 $R_{eff}$ to be the upper limit for bound systems (\citealt{Kartha2014}, \citealt{Forbes2017}, \citealt{Caso2019}). In order to obtain a nearly equal number of host-galaxy/intra-cluster sources, we considered as intra-cluster GCs those lying more than 6 $R_{eff}$ from the nearest galaxy. In this way, we found 86 host-galaxy and 82 intra-cluster GC-LMXBs. Coincidentally for NGC1399, this limit roughly corresponds to the separation between the central region covered by \textit{HST} data \citep{Paolillo2011} and the outer cluster region that has only been studied with ground-based data so far. Due to projection effects, some intra-cluster sources will be included in the host-galaxy sample. However, our conservative choice should ensure that most of the sources that are classified as intra-cluster objects are loosely bound to individual galaxies.  In Fig. \ref{fig:SD} we show the distribution of intra-cluster and host-galaxy GC-LMXBs in our field of view. The majority of intra-cluster sources are located in the central cluster region around NGC1399.

\begin{figure}
    \centering
     \includegraphics[width=\hsize]{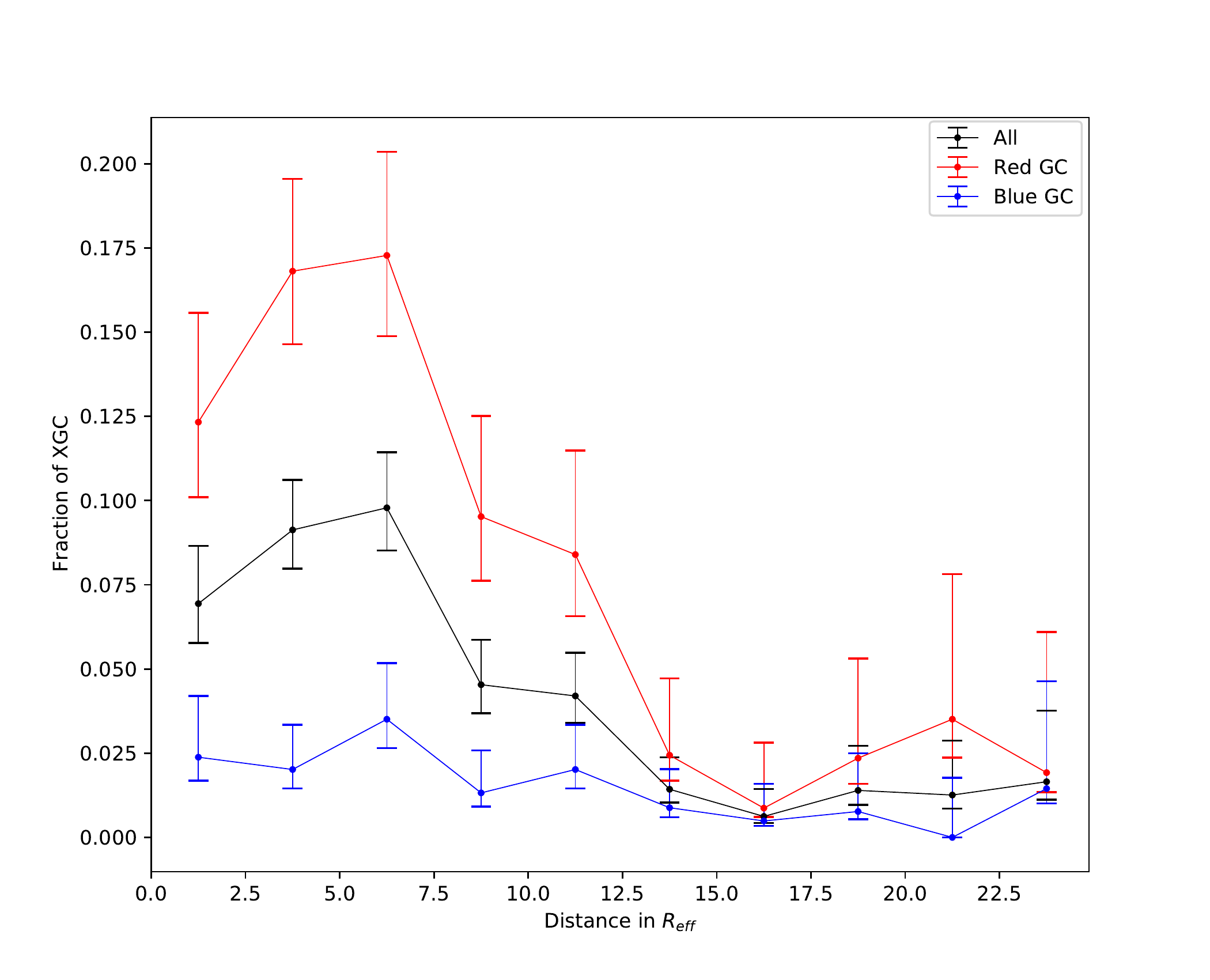}
    
    \caption{Fraction of XGC as a function of the distance in units of $R_{eff}$ for the total, red, and blue sample of GCs. The fraction is defined as the number of total, red, or blue XGC divided by the respective total number of GCs in the distance bin. The distances on the X-axis are from the nearest galaxy to the object.} 
    \label{fig:fraction}
\end{figure}

In figure \ref{fig:fraction} we show the fraction of GC hosting LMXBs (XGC) as a function of the distance from the nearest galaxy in terms of $R_{eff}$ for the entire GC sample, as well as red and blue GC subsamples. These fractions are normalized to the number of all, red, and blue GCs  in the corresponding distance bin. We refer to this as  $f_{XGC}$, $f_{rXGC}$, and $f_{bXGC}$  for the total, red, and blue fraction of XGCs, respectively. We selected only GCs with a magnitude $g<23.5$ (the 90\% completeness limit of the GC catalogue, see Fig. \ref{fig:g_lfunc}) in order to minimize the radial dependence on the optical completeness. We further verified that the same trends were confirmed when only LMXBs above the median sensitivity limit $L_x>1.7\times 10^{38}$ $erg$ $s^{-1}$ of \citet{Jin2019} were selected (corresponding to a photon flux of $1.5\times 10^{-6}$ ph s$^{-1}$ cm$^{-2}$ in their Fig.2). In this way, we also minimized the dependence on the variable X-ray completeness across the FoV. 

Figure \ref{fig:fraction} shows that the likelihood of a red GC to host an LMXB decreases with galactocentric distance, but remains approximately constant for the blue GC population. We find the host-galaxy fraction of red XGC $f_{rXGC}=0.154\pm0.017$ to be significantly different from the respective intra-cluster value $f_{rXGC}=0.062\pm0.008$. On the other hand, the host-galaxy fraction of blue XGCs $f_{bXGC}=0.021\pm0.006$ appears consistent within the errors but is still higher than that of the intra-cluster counterparts $f_{bXGC}=0.013\pm0.003$. Because of the normalization we adopted, this result is independent of the known different radial distribution of red and blue GCs. The former follow the galaxy light more closely than the latter (which represent the dominant intra-cluster GC population). Thus, the clustering of LMXBs around bright galaxies is  due to the combined effects of the steeper profile of the red GC population and the increased likelihood to host LMXBs close to the galaxy centre. Previous studies targeting nearby elliptical galaxies on spatial scales comparable to ours, such as \citet{Kim2006}, show that $f_{rXGC}$ can range from 2.7\% up to 13\% from one galaxy to the next (the latter value refers to NGC1399), while $f_{bXGC}$ remain relatively constant at $\sim$ 2\%, except for NGC1399, where it reaches 5.8\%.  We thus find a strong agreement between our host-galaxy $f_{rXGC}$ and the measurement carried out by \cite{Kim2006} on NGC1399, suggesting that our sample of host-galaxy red XGC is dominated by the population associated with the central cluster galaxy. In the case of blue XGCs, the fraction we measure is closer to the average, suggesting that blue GCs are more related to the overall cluster environment than to a single galaxy, as discussed in more detail below.

\begin{figure}
    \centering
     \includegraphics[width=\hsize]{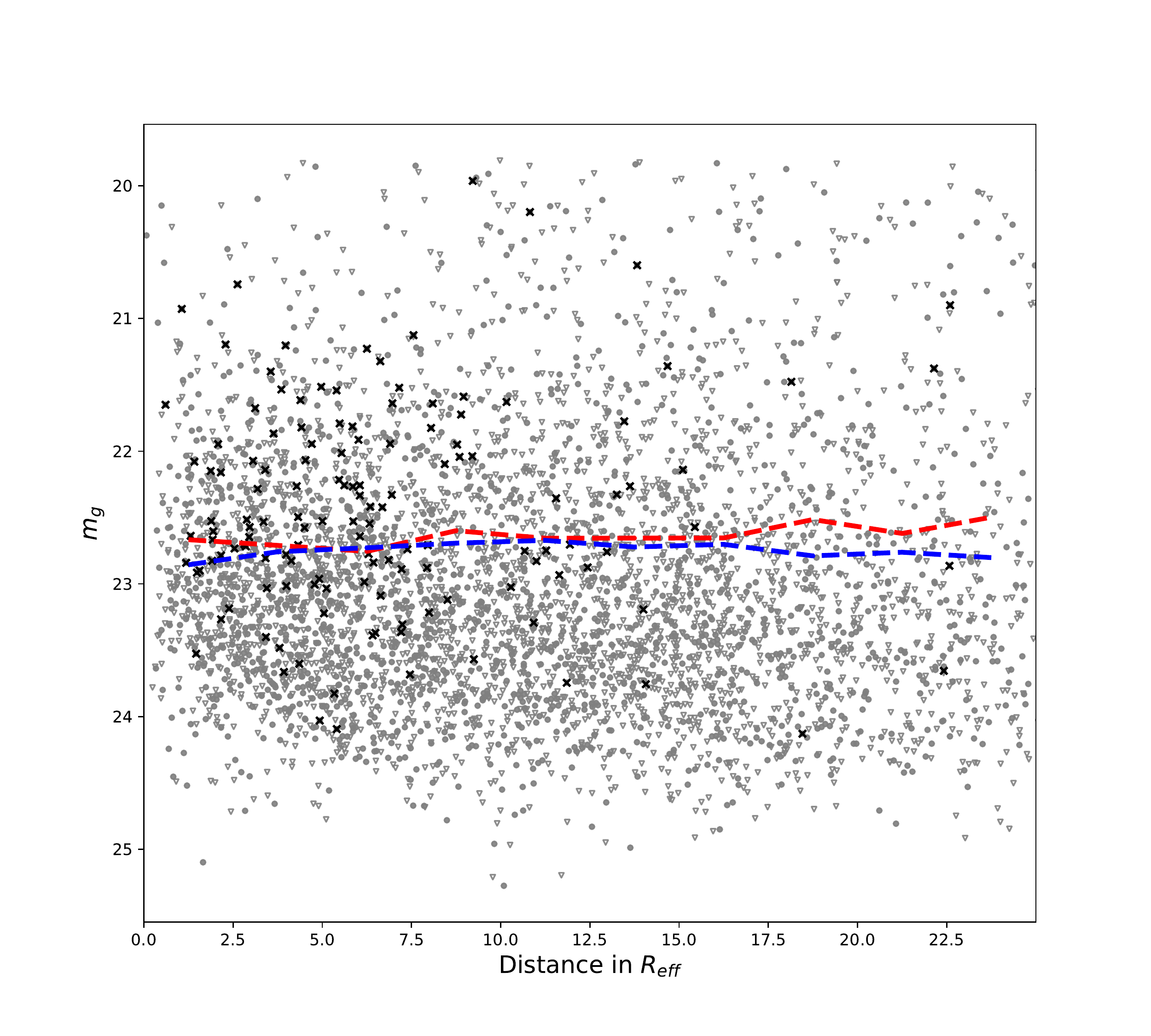}
    
    \caption{Magnitude in \textit{g} band as a function of the distance in terms of $R_{eff}$. Grey circles and triangles represent red and blue GCs, respectively. Black crosses are GCs that host an LMXB. Dashed lines represent the mean g magnitude of the GC in each distance range for the red and blue sample. The dashed lines show a computer cut of the g magnitude at g<23.5.}
    \label{fig:meanmag}
\end{figure}

Because the formation of LMXBs is strongly influenced by the luminosity of the host GC, the drop in the fraction of GC-LMXB could be associated with an average decrease in the GC luminosity towards the outer regions. In Fig. \ref{fig:meanmag} we show the GC \textit{g} -band magnitude as a function of galactocentric distance. The dashed blue and red lines represent the mean magnitude for the blue and red sample, respectively. We can observe no clear decrease in the mean magnitude towards the outer regions, which might justify the drop of the GC-LMXB fraction. This might suggest that the LMXB formation is favoured in the proximity of galaxies, or that it might also depend on other factors, such as the environment of the host galaxy.




\section{Properties of GC-LMXBs}

\subsection{Optical properties of the host GCs}
\begin{figure}[]
    \centering
     \includegraphics[width=\hsize]{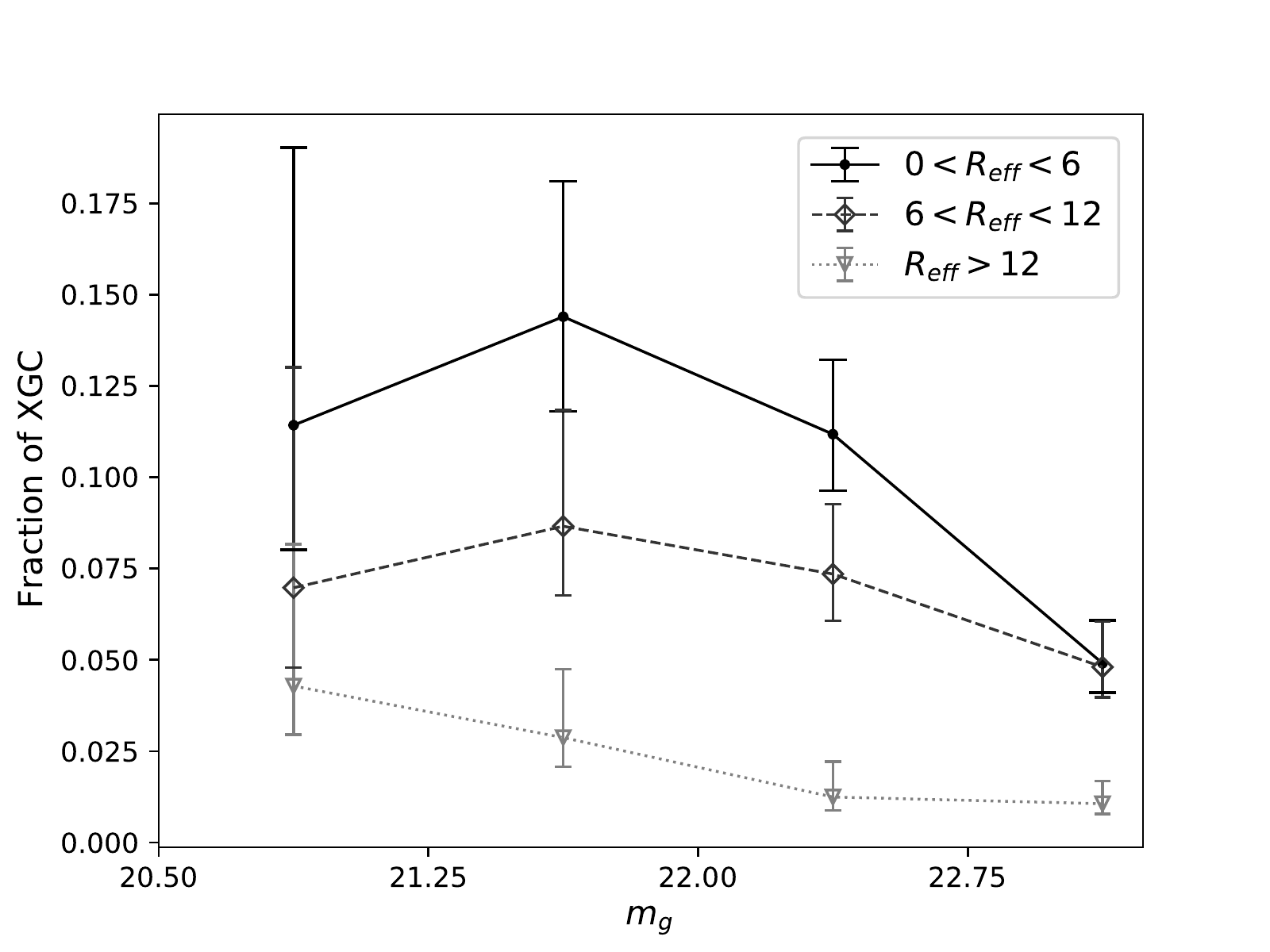}
    
    \caption{Fraction of GCs hosting LMXBs as a function of luminosity. The different types of lines correspond to different galactocentric distances. The magnitude range is restricted to $g \leq 23.5$. The fraction is defined as the number of XGCs divided by the number of GCs in the magnitude bin. The result for the inner region agrees with the results obtained in \cite{Paolillo2011}. These authors covered almost the same FoV with \textit{HST}. }
    \label{fig:frac_mag_distance}
\end{figure}

\begin{figure}[]
    \centering
    \includegraphics[width=\hsize]{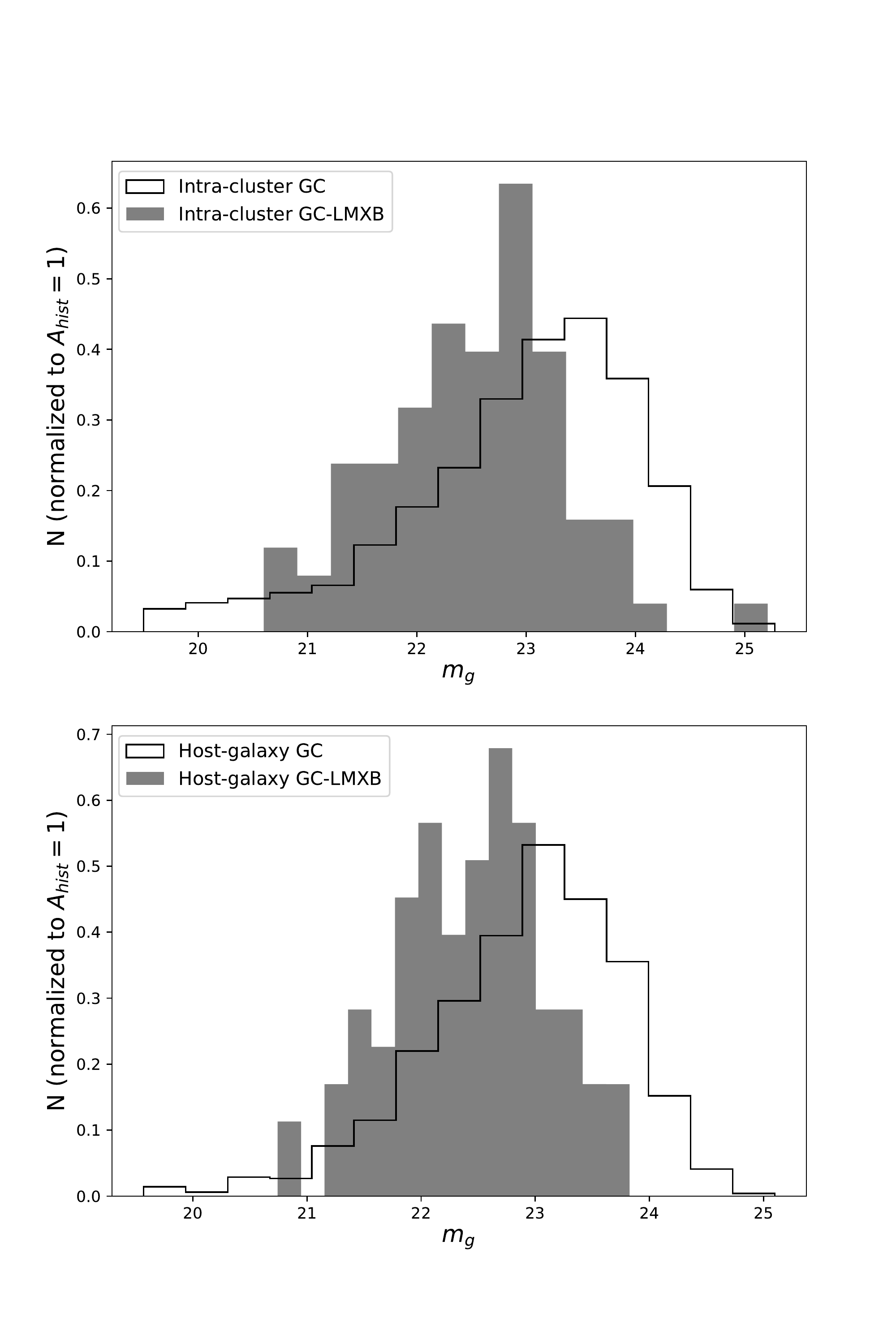}
    \caption{Magnitude distribution of GCs hosting (grey shaded) and GCs without (black line) LMXBs in intra-cluster space (upper panel) and in the host-galaxy environment (lower panel).  }
    \label{fig:mag_dist}
\end{figure}

Because most of previous works that focused on the GC-LMXB connection were restricted to the innermost regions of the galaxies, we verified whether the intra-cluster sample possesses the same properties as its host-galaxy counterpart. As a starting point, we investigated the dependence of the LMXB formation on the host GC luminosity. In Fig. \ref{fig:frac_mag_distance} we present the fraction of GC-LMXB as a function of the GC apparent magnitude in the g band for three different galactocentric distances. We note a declining trend in the host-galaxy sample ($R_{eff} \leq 6$) as well as in the intra-cluster sample ($R_{eff}>6$), suggesting that LMXBs tend to form in bright GCs in both environments. Fig. \ref{fig:mag_dist} (upper panel) confirms this result by comparing the magnitude distribution of GCs and GC-LMXBs for host-galaxy and intra-cluster sources. In both galactic and intra-cluster cases, a Kolmogorov-Smirnov (K-S) test confirms the difference between the distributions of GCs that hosting an LMXB or are without one at the 99.9\% confidence level.

\begin{figure}[]
    \centering
    \includegraphics[width=\hsize]{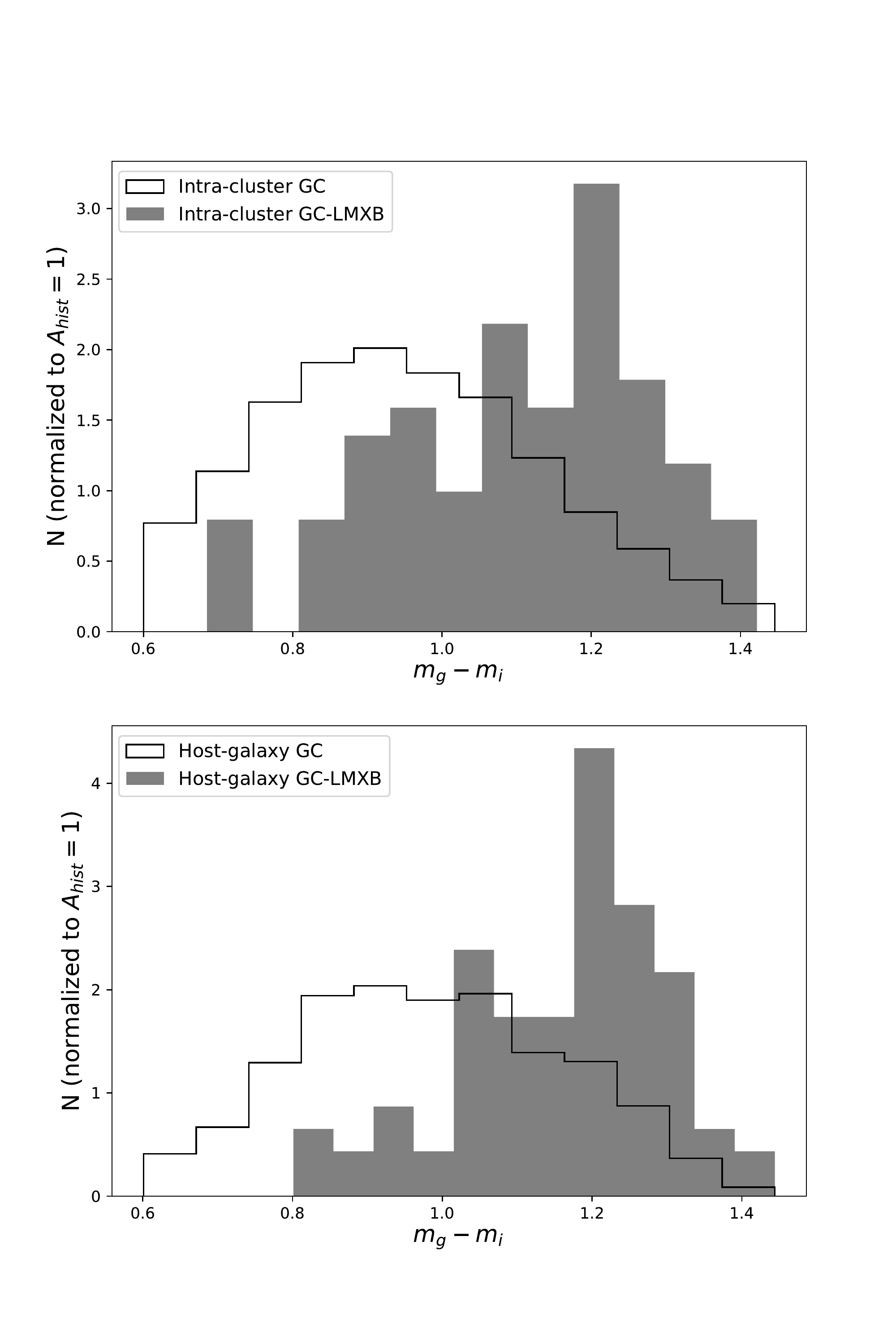}
    \caption{Colour distribution of GCs hosting (grey shaded) and GCs without (black line) LMXBs in intra-cluster space (upper panel) and in the host-galaxy environment (lower panel).  }
    \label{fig:col_dist}
\end{figure}

The LMXB formation efficiency was also shown to be dependent on the colour of the host GC (see Sec.\ref{sec:Introduction}), with a greater tendency for LMXBs to form in red GCs. To verify whether this trend also holds for intra-cluster GCs, we present in Fig. \ref{fig:col_dist} the \textit{g-i} colour distribution of intra-cluster and galactic GCs. Again, a K-S test confirms the difference in colour between GCs with and without an LMXB at the 99.9\% level in both environments.

\subsection{X-ray properties of the GC-LMXB}
\label{sec:4.2}
\begin{figure}[]
    \centering
    \includegraphics[width=\hsize]{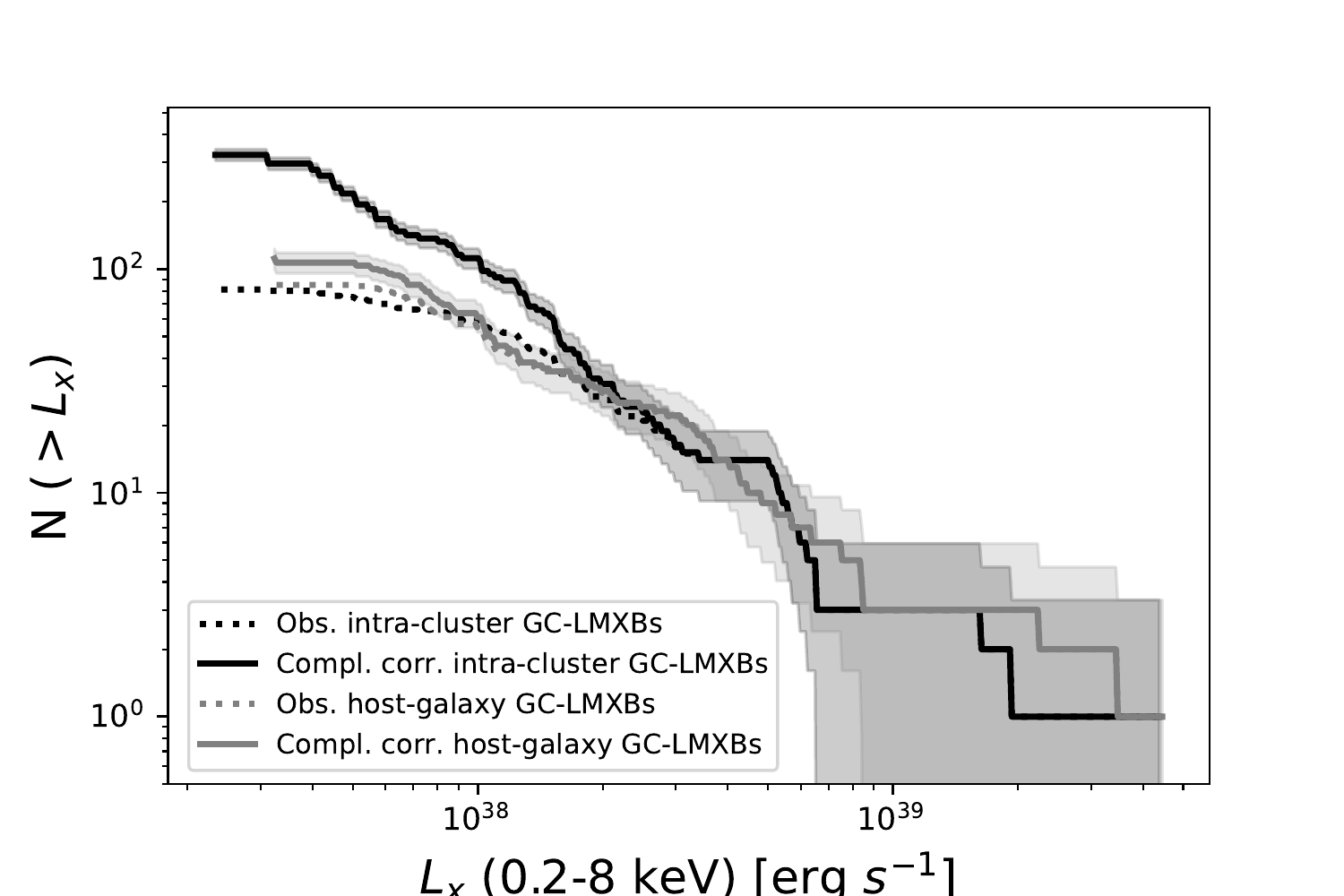}
    \caption{ Cumulative X-ray luminosity function for the host-galaxy and intra-cluster sample of GC-LMXBs. The solid lines represent the completeness-corrected LFs, and the dotted lines show the observed LFs. The shaded region represents the $1\sigma$ error.}
    \label{fig:Lf}
\end{figure}

Previous works (\citealt{Jordan2004}, \citealt{Kim2004}, \citealt{Paolillo2011}) showed that the X-ray properties of LMXBs, such as their X-ray luminosity function, do not depend on the properties of the stellar environment in which they form, such as the colour, magnitude, or density of the host GC. In this section we try to understand if these statements are valid for the intra-cluster population of GC-LMXBs. In order to correct the completeness of our X-ray source catalogue for the variable detection limit across the FoV due to hot gas emission near the brightest galaxies, the variable PSF with off-axis angle, and the different exposure time across the X-ray mosaic, we used the X-ray sensitivity map produced by \cite{Jin2019} to weight X-ray sources according to the fraction of the GCs in which they could have been detected. In Figure \ref{fig:Lf} we show the observed and completeness-corrected X-ray luminosity function of the intra-cluster and host-galaxy samples of GC-LMXBs. We found that the completeness-corrected intra-cluster and host-galaxy populations follow a single power law down to $L_x\simeq 8.5\times 10^{37}$ erg s$^{-1}$\footnote{This limit yields the best  power-law fit according to the algorithm of \cite{Newman2005} as implemented in the R routine {\it powerlaw.R}.}. The intra-cluster sample is more affected by incompleteness because the sensitivity at the edges of the FoV is lower. Fitting the completeness-corrected distributions, we obtain an LF slope of $\alpha=2.04\pm0.13$ for the host-galaxy sample (in agreement with the slopes found in previous works for GC-LMXBs belonging to galaxies, e.g.  \citealt{Kim2006}, \citealt{Paolillo2011}, \citealt{Jin2019}), and $\alpha=2.44\pm0.13$ for the intra-cluster sample. This is considerably steeper and consistent with the slope found for field LMXBs in \cite{Jin2019} ($\alpha=2.30\pm0.12$) and in previous works (\citealt{Paolillo2011}). A K-S test confirms the difference of the two completeness-corrected LFs at the 99\% confidence level. 

\begin{figure}[]
    \centering
    \includegraphics[width=\hsize]{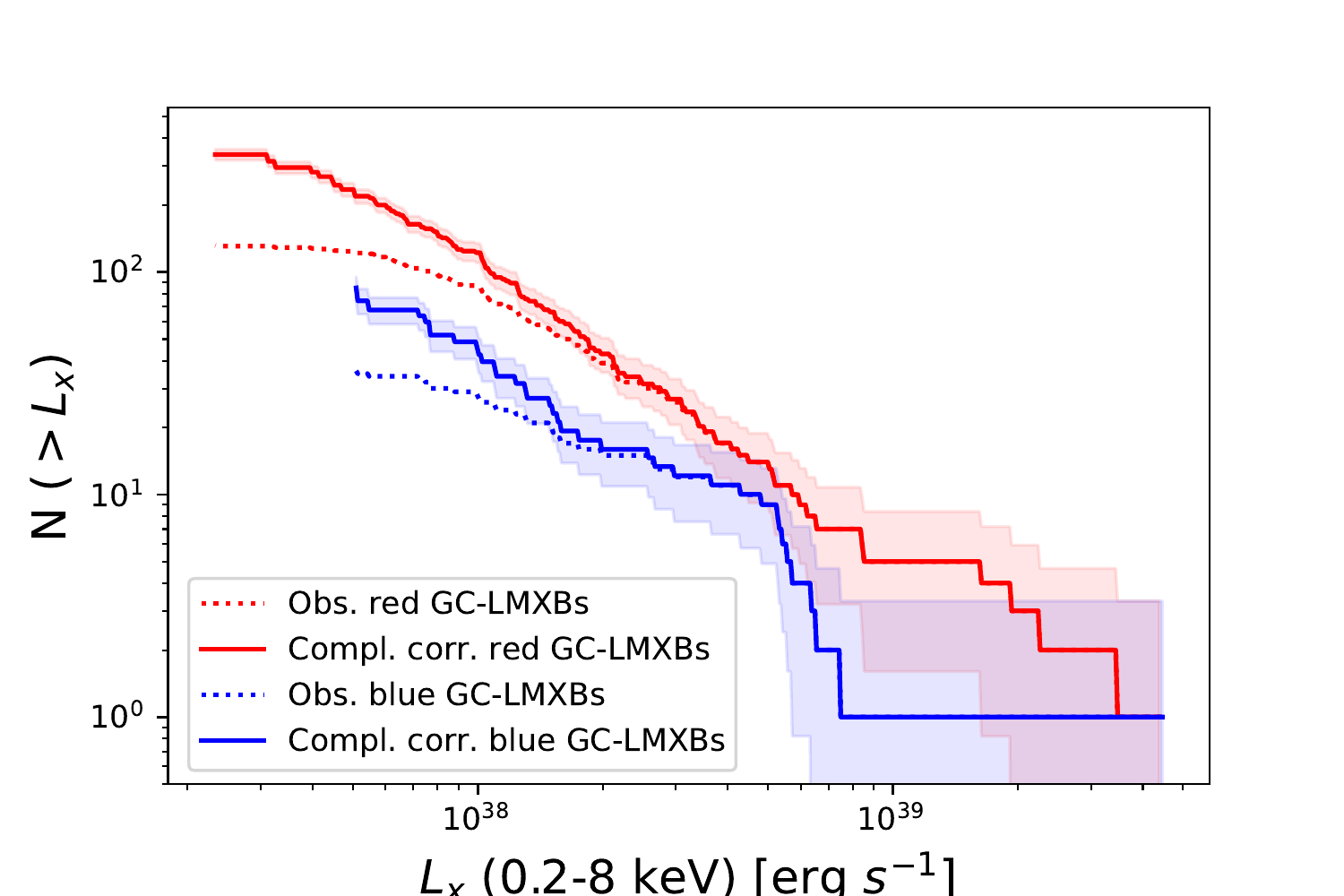}
    \includegraphics[width=\hsize]{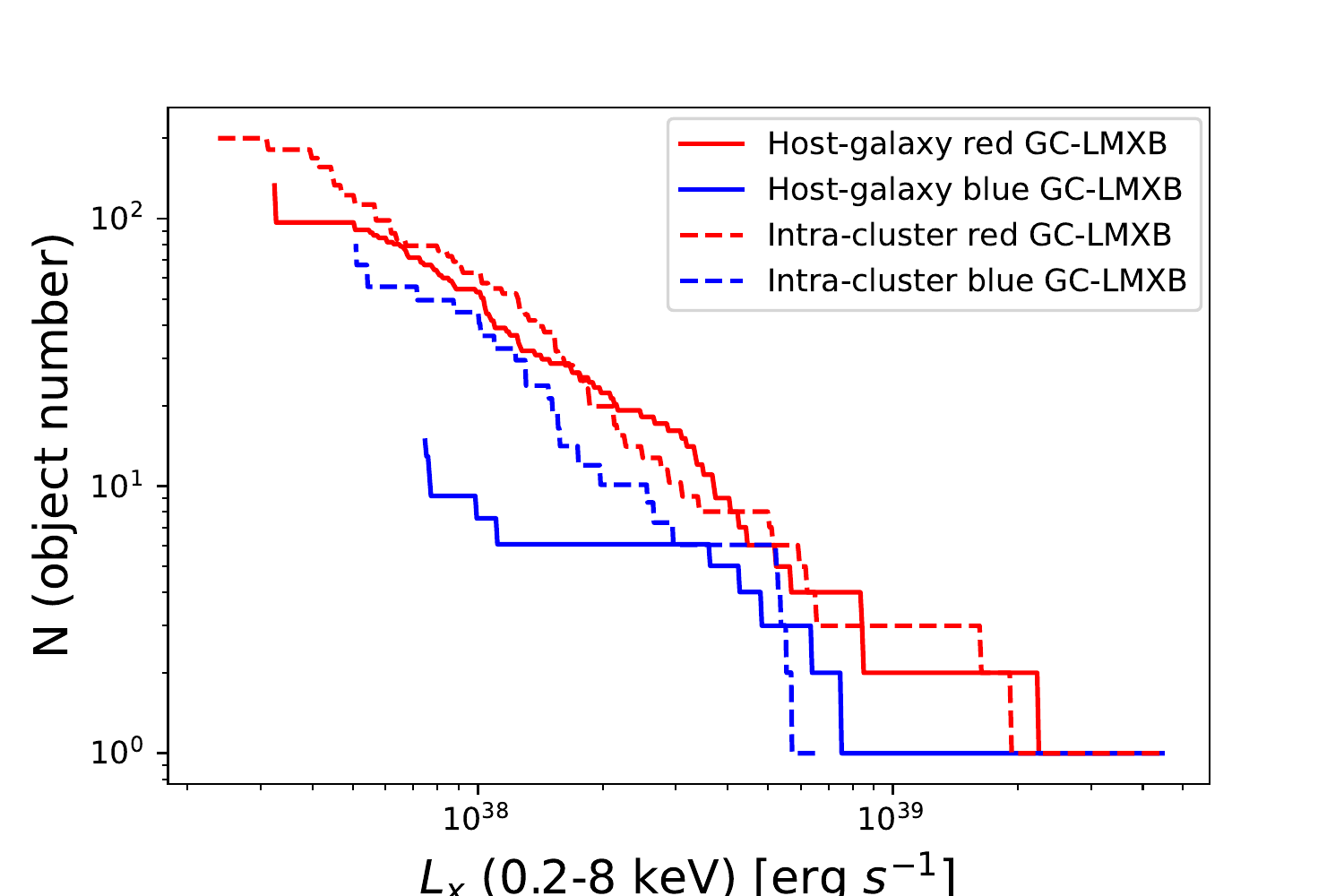}
   
    \caption{Cumulative X-ray luminosity functions comparison. \textit{Upper panel}. Red and blue sample of GC-LMXB. Lines and shaded areas are represented as in Figure \ref{fig:Lf}. \textit{Lower panel}. Completeness-corrected luminosity functions for red and blue intra-cluster and host-galaxy sources separately.    
    }
    \label{fig:red_blue_LF}
\end{figure}

\begin{figure}[]
    \centering
    \includegraphics[width=\hsize]{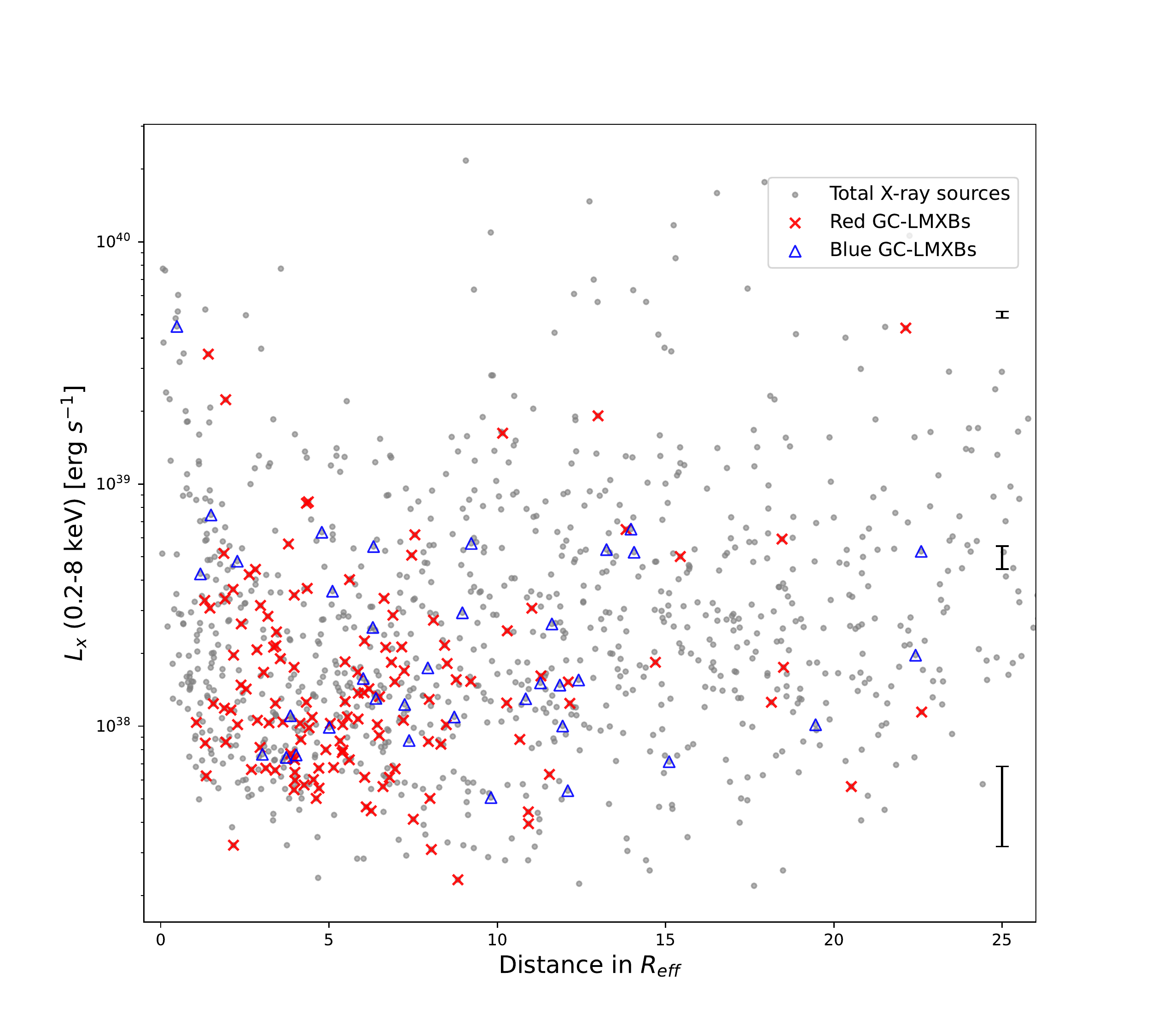}
    \caption{X-ray luminosity as a function of distance from each galaxy. Grey dots represent the whole sample of X-ray sources. Blue triangles and red crosses represent the respective GC-LMXBs. The mean errors in each X-ray luminosity order of magnitude are shown in black.  }
    \label{fig:lxvsdist}
\end{figure}

Then, we studied the $L_x$ distribution by dividing the sample of GC-LMXBs according to the host GC colour (Fig. \ref{fig:red_blue_LF}). The completeness-corrected red GC-LMXB follows a power-law distribution with a slope $\alpha=2.37 \pm 0.12$ down to $L_x\simeq 8.5\times 10^{37}$ erg s$^{-1}$ . A possible lack of bright LMXBs is observed in the blue systems, which is found to be significant at the $\sim 2\sigma$ (97\%) level according to a Poisson statistics. 
Fitting the entire sample of blue GC-LMXBs down to $L_x\simeq 8.5\times 10^{37}$ erg s$^{-1}$ yields $\alpha=2.21 \pm 0.18$.
To further investigate this trend, we separated the red and blue sample into intra-cluster and host-galaxy sources (lower panel of Fig. \ref{fig:red_blue_LF}). The population of blue intra-cluster GCs does not host LMXBs above $\sim 6\times 10^{38}$ $erg/s$, although again the number of object is small for strong conclusions. The majority of objects above $10^{39}$ $erg / s$ resides in red GCs possibly because multiple LMXBs reside in a single GC because it is easier to form LMXBs in these systems. However, previous studies often detected variability in systems like this, which argues against this possibility. On the other hand, this high luminosity break has already been observed in literature (\citealt{Fabbiano2006}, and reference therein) and may reflect the transition between the most massive neutron stars and low-mass black hole systems, suggesting that blue environments are less likely to form binary systems with massive black holes. We point out, however, that considering the small statistics at high $L_x$, a small contamination by background AGNs randomly matched with GCs could be enough to cause the difference (see below). To verify that this is not linked to the different host GC distribution, we show in Fig. \ref{fig:lxvsdist} the correlation between $L_x$ and the colour/galactocentric distance of the GCs  for the red and blue sample of GC-LMXB. While red GC-LMXB are more centrally concentrated, as discussed before, the brightest sources with $L_X>10^{39}$ erg s$^{-1}$ 
are uniformly distributed throughout the whole distance range. 
The small number of sources with $L_X>10^{39}$ erg s$^{-1}$ does not allow us to draw definitive conclusions, and therefore prevents us from excluding the effect of a background AGN contamination.

\begin{figure}[]
    \centering
    
    \includegraphics[width=\hsize]{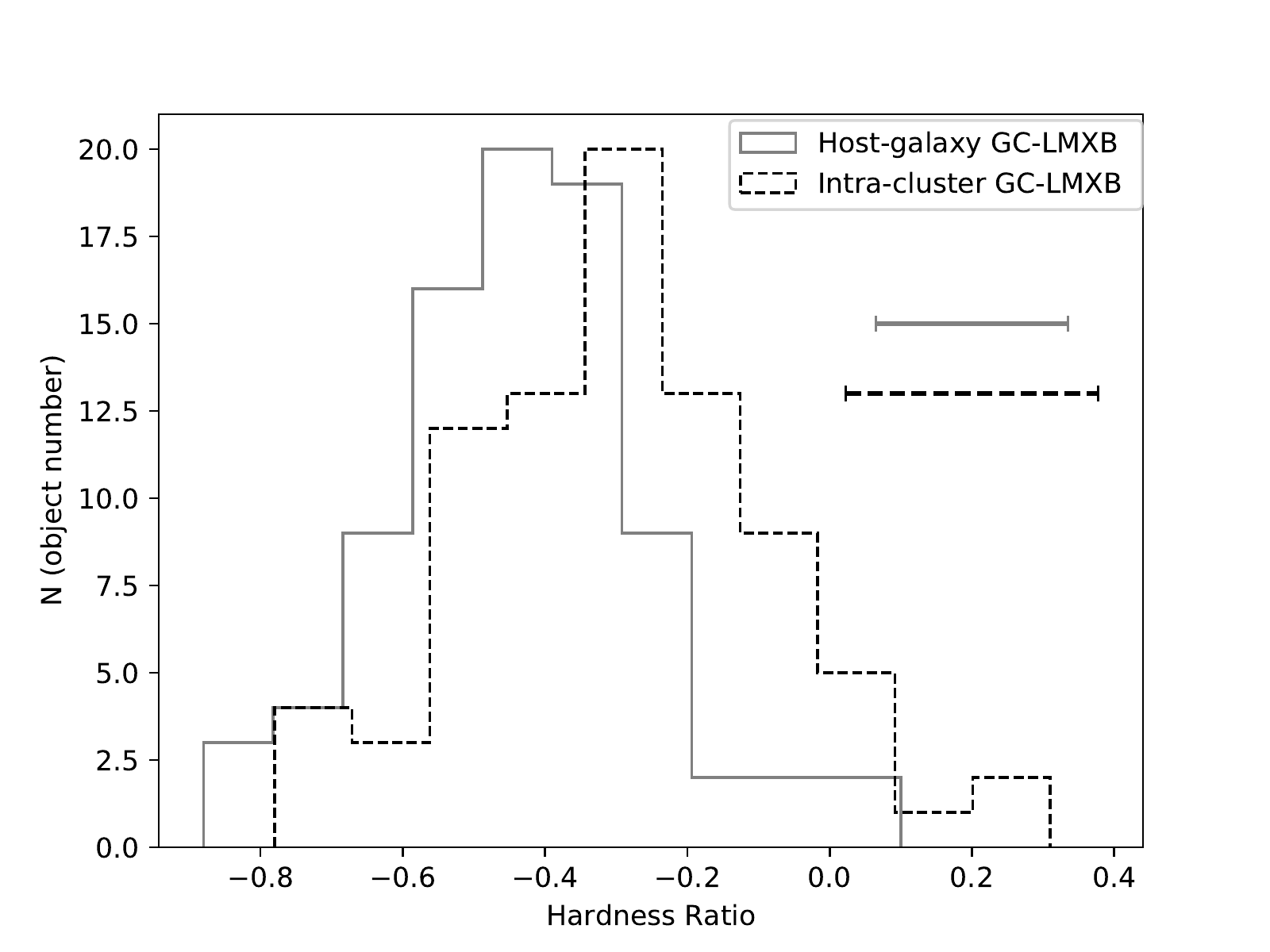}
    \caption{Hardness ratios between host-galaxy (grey) and intra-cluster (black) GC-LMXBs. The error bar is the median value of the errors associated with the HR of the sources.} 
    \label{fig:HR}
\end{figure}

We further investigated the spectral properties of the LMXB population through their hardness-ratio (HR), defined as $HR=\frac{S_{Hard}-S_{Soft}}{S_{Hard}+S_{Soft}}$, with $S_{hard}$ the photon flux in the hard band (2-8 KeV) and $S_{soft}$ the photon flux in the soft band (0.5-2 Kev). In Fig. \ref{fig:HR} we present the HR distribution of host-galaxy and intra-cluster sources. It is readily apparent that intra-cluster GC-LMXBs have harder spectra. A K-S test confirms that the two samples are drawn from different distributions. To understand this trend, we considered whether this systematic difference is related to the LMXB population itself or to their host GCs because we know from Sec.\ref{optical_data} and from previous works that the GC population becomes increasingly blue with increasing galactocentric distance \citep{Jordan2006, D'Abrusco2016, Cantiello2018, Cantiello2020}. 

\begin{figure}[]
   
    \includegraphics[width=\hsize]{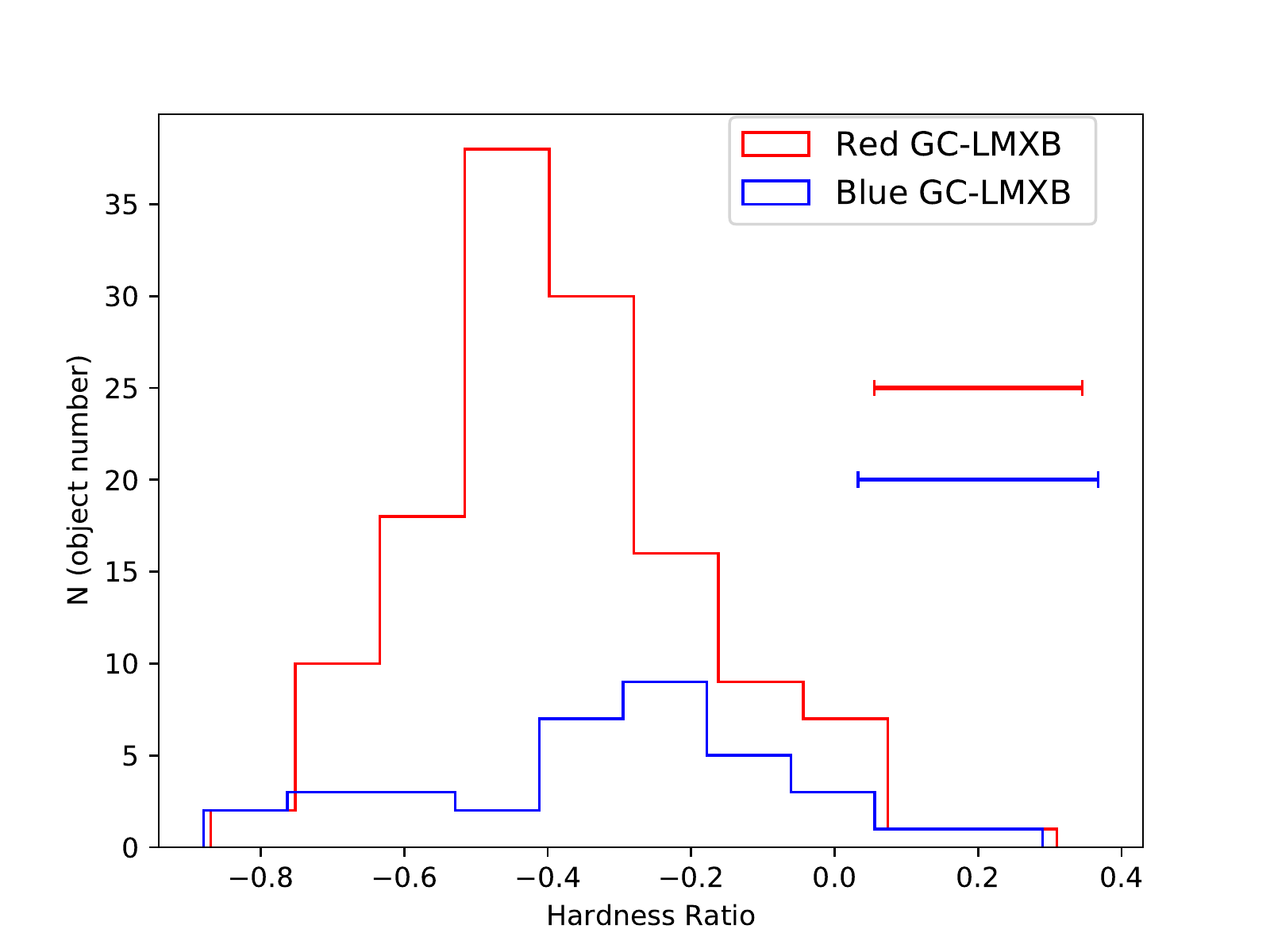}
    \caption{Hardness ratios between red (grey) and blue (black) GC-LMXBs. The error bar is the median value of the errors associated with the HR of the sources.} 
    \label{fig:HRrb}
\end{figure}

In our case, 694 blue GCs (containing 10 LMXBs, i.e. 1.4\% of the population) and 618 red GCs (72 LMXBs, i.e. 11.6\% of the population) are associated with the main cluster galaxies, while the intra-cluster sample contains 2400 blue GCs (26  LMXBs, i.e. 1.1\% of the population) against  1467 red (60  LMXBs, i.e. 4.1\% of the population). The number of blue GCs therefore increases in the outer regions. As shown in Fig. \ref{fig:GMMdifference}, we also observe a  shift in the average value of the red distribution towards bluer colours, however. 
In Fig. \ref{fig:HRrb} we show that LMXBs in blue GCs seem to have a harder spectrum than those in red GCs; the result is confirmed when red and blue GC-LMXBs are compared restricted to the intra-cluster sample alone\footnote{Comparing red and blue GC-LMXBs restricted to the host-galaxy sample alone does not allow us to draw definitive conclusions because there are too few blue GC-LMXBs.}.

To explain this result, we explored three possibilities: 1) Intra-cluster sources are more heavily contaminated by harder-backgound sources than the host-galaxy population. 2) The difference is due to uncertainties in the X-ray spectral response correction, which in turn is due to the combination of the multiple \textit{Chandra} exposures with different off-axis angles and the presence of diffuse gas, which may affect the background estimate. Finally, 3) the difference is real.

To address the first possibility, we estimated the number of expected contaminants in our FoV. To this end, we used the COSMOS  catalogues by \citet{Civano2016} and \citet{Laigle2016} to select X-ray sources within the same X-ray flux limit as was used for the catalogue of \cite{Jin2019}, and with optical counterparts obeying the same selection criteria in terms of limiting optical magnitude, colours, and concentration index as we adopted in Sec.\ref{optical_data}. We predict a contamination of $ \text{about ten}$ background X-ray sources with optical counterparts over our FoV. In addition, we expect three random superpositions of background sources without optical counterparts, as discussed in Sec.\ref{Sec:3}. These contaminants do have harder average HR than the bulk of GC-LMXB sources, but they do not account for the entire excess observed in our sample. When we removed the ten hardest sources from the sample of intra-cluster GC-LMXBs, the result was unchanged.

\begin{figure*}[]
    \centering
    
    \includegraphics[width=0.5\hsize]{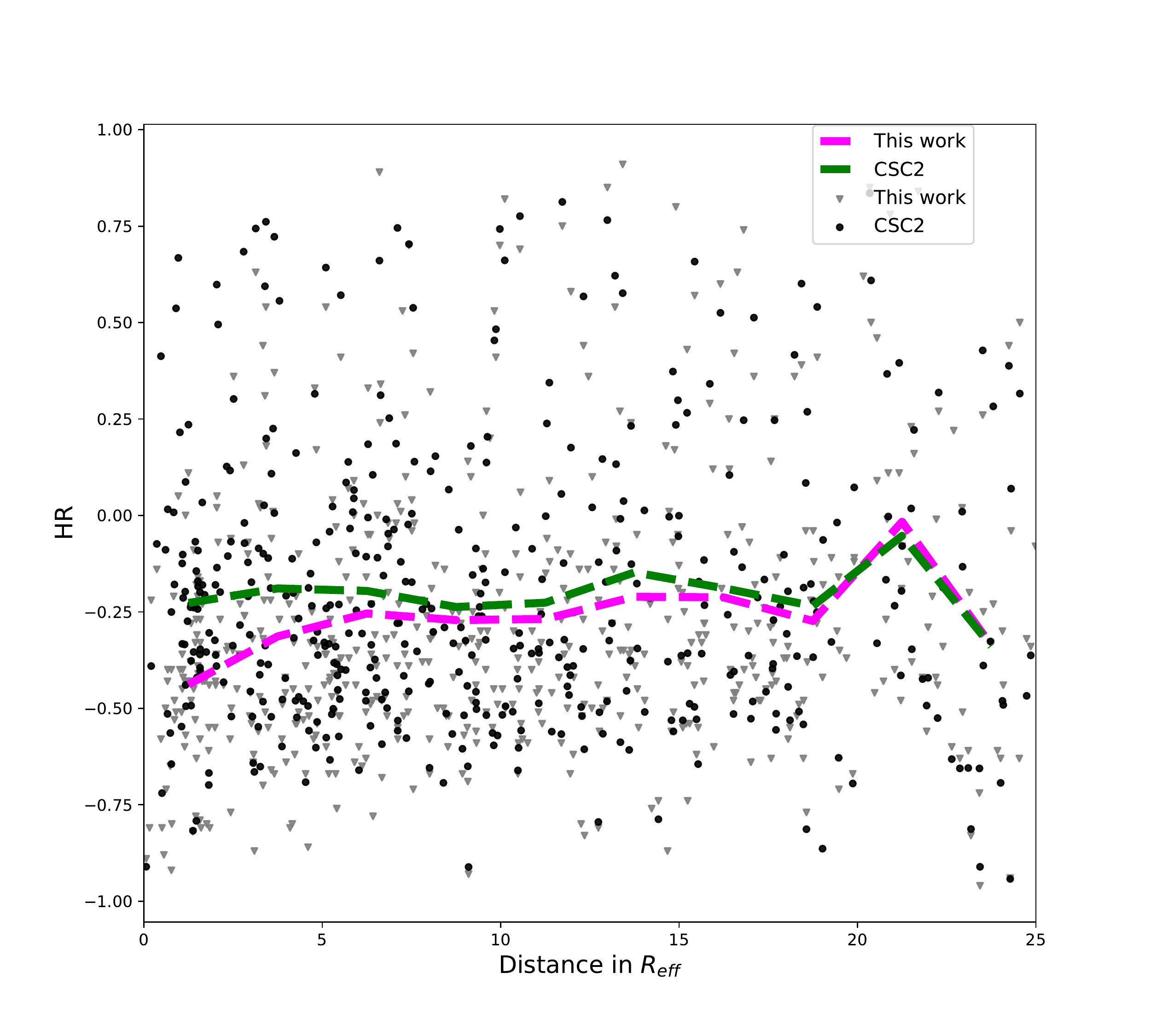}%
    \includegraphics[width=0.5\hsize]{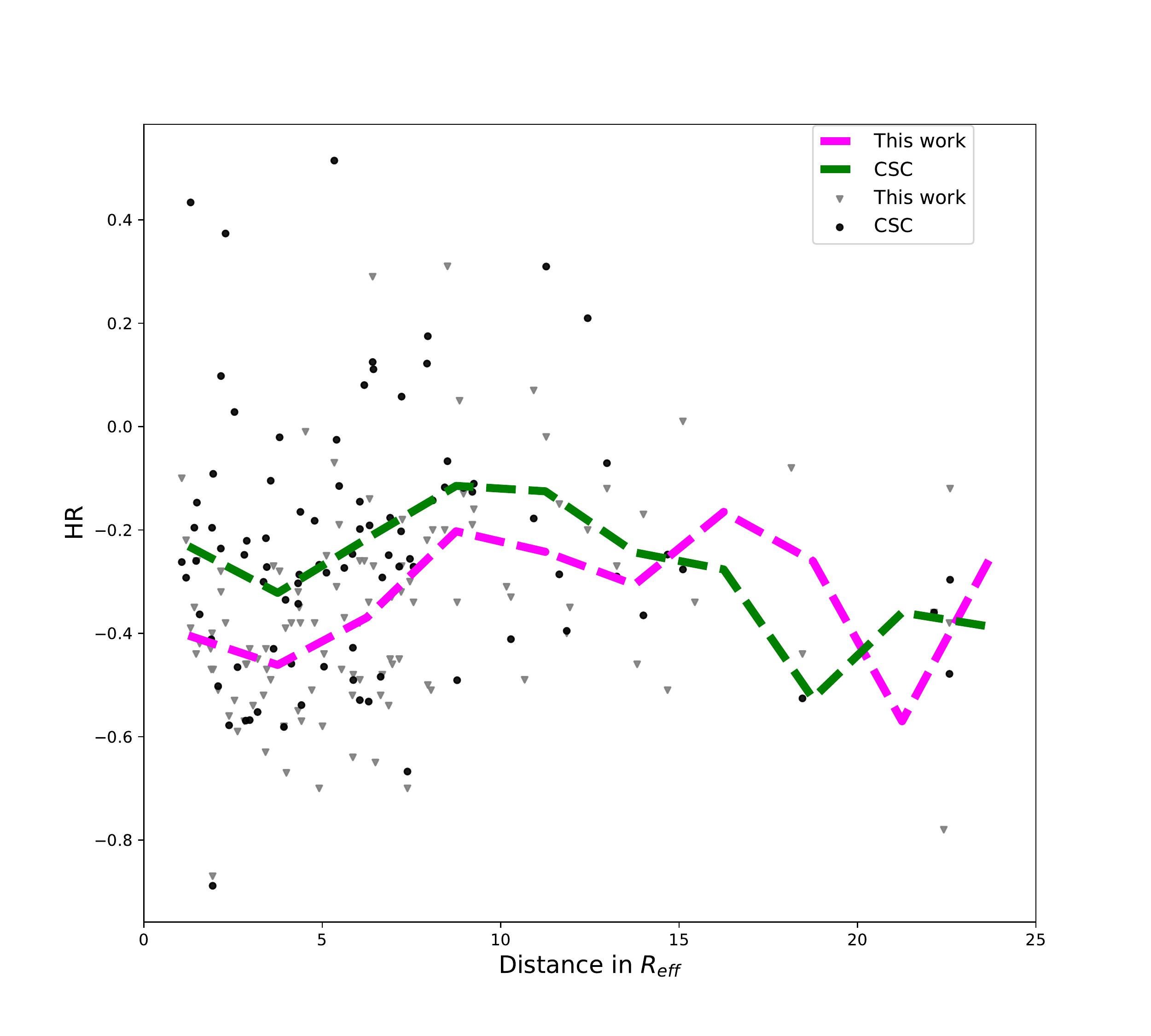}
    \caption{ Hardness ratio as a function of the distance for field X-ray objects (left panel) and GC-LMXBs (right panel) with counterparts in the CSC catalogue. Black dots represent the HR estimates in the CSC catalogue, and grey triangles show the HR we used. Dashed green and magenta lines again represent the mean value in each distance bin. }
    \label{fig:HRCSC333}
\end{figure*}

Several factors can affect the derived hardness ratios in the spectral response, including different off-axis angles, detectors (ACIS-I versus ACIS-S), and observing epoch (due to the efficiency degradation of ACIS at low energies with time). We point out, however, that \cite{Jin2019} only used ACIS I0, I1, I2, I3, S2, and S3 chips in order to ensure an optimal source sensitivity. In addition, most intra-cluster sources are still located in the central region of the cluster, although at a greater distance from the main galaxies, and they are covered by multiple observations  spanning several years (see Table 1 in \citealt{Jin2019}). These properties should reduce the potential systematic effects in the X-ray catalogue. To evaluate how the catalogue properties might affect our results, we compared the HR estimates used in this work with the estimate reported in the CSC  catalogue. In Fig. \ref{fig:HRCSC333} we show the HR as a function of the distance from the nearest galaxy for X-ray sources with a counterpart in the CSC master catalogue. We note a difference in the mean HR, mainly driven by {\it host-galaxy} sources, both considering GC-LMXBs and field X-ray sources. The higher HR in the CSC is higher on average than the HRs reported in the catalogue used in this work. This suggests that the spectral corrections differ between the two catalogues, possibly due to the different data reduction strategy, source extraction method (see Sec.\ref{sec:2.2}), and the treatment of the contribution of the central diffuse emission.

Concerning the possibility that the observed difference might be real, it is known that the colour mainly depends on metallicity in old GCs, where metal-rich GCs are significantly redder (\citealt{Cantiello2018}). This might suggest a relation between the spectral properties of LMXBs and the metallicity of the host GC. One model that discusses such a trend was presented by \cite{Maccarone2004}, who proposed irradiation-induced winds in metal-poor stars that would cause absorption, mainly of the soft part of the X-ray spectra, which would produce higher HR values. The authors reported this trend for NGC4472 (\citealt{Maccarone2003}). In this case Fornax, would be the only other system in which this effect is directly observed so far. 

\section{Summary and conclusions}

We performed an analysis of the properties of LMXBs residing in the population of intra-cluster GCs in the Fornax cluster. The main goal of this work was to study the GC-LMXB connection within the core of the Fornax cluster and its dependence on the environment in terms of galactocentric distance and host GC. For this purpose, we used optical photometry of 5178 candidate GCs from the Fornax Deep Survey \citep{Cantiello2020}. In order to study the possible differences between LMXBs that formed in different GC environments, we performed a \textit{Gaussian Mixture Model} fit on the g-i colour distribution, finding a separation threshold of $g-i \sim 1.00$ between red and blue that agrees with the threshold reported in previous works (\citealt{D'Abrusco2016}, \citealt{Cantiello2020}). We obtained 2085 red GCs and 3093 blue GCs. 
We further observe a tendency in the average $g-i$ colour of becoming bluer with increasing galactocentric distance. Because the GC colours mainly depend on metallicity, this result suggests that the intra-cluster GCs have a lower metallicity on average. 

The X-ray data employed in this work were extracted from archival \textit{Chandra} observations. The source detection procedure together with the extraction of photometry  was performed by \cite{Jin2019}.
In order to study the connection between LMXBs and GCs, we performed a cross-match between the optical and the X-ray catalogues. In this way, we identified 168 X-ray sources that are positionally coincident with GCs. We divided this population into \textit{host-galaxy} and \textit{intra-cluster} objects based on their projected distance from the nearest galaxy. We considered as \textit{intra-cluster} objects the GC-LMXBs within more than 6 $R_{eff}$ from the nearest galaxy and found 82 intra-cluster and 86 host-galaxy GC-LMXBs. Furthermore, considering the colour division performed for the GC sample, we found 36 LMXBs to be associated with blue GCs and 132  LMXBs to be associated with red GCs. 

We find the fraction of GC-LMXBs to be dependent on the galactocentric distance; this effect is particularly evident for the red population. Because the GC magnitude seems to be independent of the distance from the galaxies, we conclude that this result may suggest that the LMXBs formation channel in GCs may also depend on the host-galaxy environment. In the past, the evidence for a dependence of the LMXB formation likelihood on the local galaxy environment has been debated (see e.g. \citealt{Kim2006} and \citealt{Paolillo2011} for an opposing view), but these studies were essentially limited to the inner region of galaxies within a few $R_{eff}$. Even in this work, there is little evidence for a dependence like this within $\sim 6~  R_{eff}$ , which is visible only when the analysis is extended to the intra-cluster population. If this is confirmed, the enhanced LMXB formation rate in red GCs might be related to their orbital parameters, as suggested by \cite{Puzia2014}, for instance, leading to a stronger influence of the external tidal field.  \cite{Webb2016} found that the sizes of red GCs in NGC1399 are consistent with more radial orbits compared to blue GCs.

We confirm that intra-cluster LMXBs tend to form in red and bright GCs, as has been found for their host-galaxy couterparts. Furthermore, we studied the X-ray properties of the intra-cluster population of GC-LMXB to test whether they are independent of the properties of the host GC as for the host-galaxy sources. We find that the completeness-corrected X-ray luminosity function of the intra-cluster population of GC-LMXBs follows a power law with a slope that is marginally consistent ($\sim 2.2\sigma$) with the slope of the host-galaxy population, and it is consistent with the slope found for field LMXBs in the literature. A Kolmogorov-Smirnov test indicates a statistically significant difference between the $LF$ of the intra-cluster and that of the host-galaxy sample. We cannot confirm any difference between the completeness-corrected LFs of the red and blue populations. We find a lack of bright LMXBs in blue GCs, however, which agrees with what has been found for host-galaxy sources and possibly indicates a lack of black hole binaries in metal-poor systems.

Finally, we observed a puzzling  difference in hardness ratio between intra-cluster and host-galaxy GC-LMXBs: the spectra of the intra-cluster sample are harder  than those of the host-galaxy sample. 
Because intra-cluster GCs are bluer on average than the host-galaxy GCs, this result might suggest a relation between HR and the colour of the host GC, and hence the metallicity. We explored different explanations for this difference, including residual systematics in the data and background contamination or a possible physical origin. We found that contamination alone seems unable to explain the observed trend. This result is still tentative, however, and a final conclusion will have to wait for a full spectral analysis of the host-galaxy and intra-cluster GC-LMXB populations.

\begin{acknowledgements}
 GR acknowledges support from the National Science Centre (UMO-2018/30/E/ST9/00082, UMO-2018/30/M/ST9/00757 and UMO-2020/38/E/ST9/00077). 
 GD acknowledges support from FONDECYT REGULAR 1200495, and ANID project Basal FB-210003. This research has made use of data obtained from the \textit{Chandra} Data Archive and the \textit{Chandra} Source catalogue, and software provided by the \textit{Chandra} X-ray Center (CXC) in the application packages CIAO and Sherpa.
 \end{acknowledgements}

%
%
\bibliographystyle{aa} 
\bibliography{main}

\end{document}